\documentclass[11pt,a4paper]{article}            
               
\usepackage[skins,theorems]{tcolorbox}
\tcbset{highlight math style={enhanced,
  colframe=red,colback=white,arc=0pt,boxrule=1pt}}
\usepackage[bookmarksopen, bookmarksnumbered, bookmarksopenlevel=2]{hyperref}
\usepackage{tikz}
\usepackage{tikz-3dplot}
\usetikzlibrary{calc}
\usetikzlibrary{decorations} %
\usepackage[UKenglish]{babel}
\usepackage[toc,page]{appendix}
\usepackage{amsmath}
\usepackage{amssymb}
\usepackage{graphicx}
\usepackage{hhline}
\usepackage[bf]{caption}
\usepackage{cite}
\usepackage[vcentermath]{youngtab}
\usepackage{geometry}
\usepackage{slashed}
\usepackage{color}
\usepackage{stackrel}
\usepackage{tikz}
\usepackage{tikz-cd} 
\usepackage{empheq}

\newcommand{\bqa}{\begin{eqnarray}}
\newcommand{\eqa}{\end{eqnarray}}

\hypersetup{
    pdftitle={},
    pdfauthor={},
    pdfsubject={}
}
\numberwithin{equation}{section}
\numberwithin{table}{section}

\makeatletter

\newcommand{\be}{\begin{equation}}
\newcommand{\ee}{\end{equation}}
\newcommand{\ba}{\begin{aligned}}
\newcommand{\ea}{\end{aligned}}

\newcommand{\bea}{\begin{eqnarray}}
\newcommand{\eea}{\end{eqnarray}}

\newcommand*\widefbox[1]{\fbox{\hspace{2em}#1\hspace{2em}}}

\def\unit{{1\kern-.65ex {\rm l}}}
\def\1{{1\kern-.65ex {\rm l}}}

\newcount\hour \newcount\minute
\hour=\time \divide \hour by 60
\minute=\time
\def\now{%
\ifnum \hour<13
  \ifnum \hour=0 \advance \hour by 12 \number\hour:\else \number\hour:\fi%
     \ifnum \minute<10 0\fi%
     \number\minute%
\ A.M.%
\else \advance \hour by -12 \number\hour:%
  \ifnum \minute<10 0\fi%
  \number\minute%
  \ P.M.%
\fi%
}

\makeatother

\begin{document}

\baselineskip=14pt
\parskip 5pt plus 1pt

\vspace*{-1.5cm}
\begin{flushright}    
  {\small

  }
\end{flushright}

\vspace{2cm}
\begin{center}        
  {\LARGE The Green-Schwarz Mechanism and Geometric Anomaly Relations} \vspace{2mm}\\
  {\LARGE in 2d (0,2) F-theory Vacua}
\end{center}

\vspace{0.75cm}
\begin{center}        
Timo Weigand$^{1,2}$ and  Fengjun Xu$^1$
\end{center}

\vspace{0.15cm}
\begin{center}        
$^1$\emph{Institut f\"ur Theoretische Physik, Ruprecht-Karls-Universit\"at, 
\\       Philosophenweg 19, 69120,
Heidelberg, Germany \vspace{2mm}\\
           }
$^2$\emph{CERN, Theory Division, \\
             CH-1211 Geneva 23, Switzerland}
             \\[0.15cm]
          
\end{center}
\vspace{2cm}


\begin{abstract}
\noindent

We study the structure of gauge and gravitational anomalies in 2d $N=(0,2)$ theories obtained by compactification of F-theory on elliptically fibered Calabi-Yau 5-folds.
Abelian gauge anomalies, induced at 1-loop in perturbation theory, are cancelled by a generalized Green-Schwarz mechanism operating at the level of chiral scalar fields in the 2d supergravity theory. We derive closed expressions for the gravitational and the non-abelian and abelian gauge anomalies including the Green-Schwarz counterterms. These expressions  involve topological invariants of the underlying elliptic fibration and the gauge background thereon.
Cancellation of anomalies in the effective theory predicts intricate topological identities which must hold on every elliptically fibered Calabi-Yau 5-fold. We verify these relations in a non-trivial example, but their proof from a purely mathematical perspective remains as an interesting open problem.
Some of the identities we find on elliptic 5-folds are related in an intriguing way to previously studied topological identities governing the structure of anomalies in 6d $N=(1,0)$ and 4d $N=1$ theories obtained from F-theory.

\end{abstract}

\thispagestyle{empty}
\clearpage
\tableofcontents
\thispagestyle{empty}


\newpage
\setcounter{page}{1}
\section{Introduction}

Quantum anomalies considerably constrain the structure of chiral gauge theories in even dimensions. 
Chiral matter is known to induce gauge and gravitational anomalies at the 1-loop level in perturbation theory \cite{AlvarezGaume:1983ig}, which jeopardize the consistency of the gauge theory.
In the presence of tensor fields the celebrated Green-Schwarz-Sagnotti-West mechanism \cite{Green:1984sg,Green:1984bx, Sagnotti:1992qw}  can cancel such 1-loop anomalies provided the anomaly polynomial of the latter factorises suitably. A particularly interesting class of examples of such tensors are the self-dual tensor fields in $4k+2$ dimensions \cite{Witten:1996hc}.
The ramifications of  the anomaly cancellation mechanism have been investigated in great detail, most notably in the context of 6d $N=(1,0)$ supergravity theories (see e.g. \cite{Schwarz:1995zw,Kumar:2009us,Kumar:2009ac,Park:2011wv,Monnier:2017oqd} and references therein). 
A lower-dimensional analogue of these 6d $N=(1,0)$ supergravities, similar in many respects, are chiral theories in two dimensions with $N=(0,2)$ supersymmetry.
Such theories have sparked significant interest from various field theoretic perspectives, most notably concerning their RG flow to an SCFT point  \cite{Benini:2012cz,Benini:2013cda,Benini:2015bwz,Gadde:2014ppa,Couzens:2017way}, in the context of computing elliptic genera and localisation \cite{Closset:2015ohf}, or with respect to novel types of dualities \cite{Gadde:2013lxa,Jia:2014ffa}. 
Exploring the structure of anomalies of a class of 2d $N=(0,2)$ supergravities is the goal of this article.

If a supergravity theory is engineered by compactifying string theory, the consistency conditions from anomaly cancellation imply a rich set of constraints on the geometry defining the compactification. A prime example of this fruitful interplay between anomalies and geometry is provided by F-theory \cite{Vafa:1996xn,Morrison:1996na,Morrison:1996pp}. 
In this framework, 6d $N=(1,0)$ supergravities arise via compactification on elliptically fibered Calabi-Yau 3-folds.
Anomaly cancellation then translates into various highly non-trivial relations between topological invariants of the latter  \cite{Sadov:1996zm,Grassi:2000we, Grassi:2011hq, Park:2011ji, Park:2011wv}, which would be hard to guess otherwise, and some of which are even harder to prove in full generality. 
Compactification of F-theory to four dimensions on a Calabi-Yau 4-fold gives rise to an $N=1$ supersymmetric theory which is chiral  - and hence potentially anomalous - only in the presence of non-trivial gauge backgrounds.
This makes it perhaps even more intriguing that 
the same types of topological relations \cite{Bies:2017abs} are responsible for the cancellation of gauge and mixed gauge-gravitational anomalies in six and four-dimensional \cite{Cvetic:2012xn} F-theory compactifications. 
If one is able to 
establish the cancellation of anomalies directly from a physical perspective, as has been achieved recently in \cite{Corvilain:2017luj} for four-dimensional F-theory vacua, such reasoning amounts to a physics proof of a number of highly non-trivial topological relations on elliptic fibrations of complex dimension three and four.  One of the motivations for this work is to extend this list of topological identities to elliptic fibrations of higher dimension.

The 2d $(0,2)$ supergravity theories considered in this article are obtained by compatifying F-theory on an elliptically fibered Calabi-Yau 5-fold \cite{Schafer-Nameki:2016cfr, Apruzzi:2016iac}.  As we will review in section \ref{sec_FonellCY5} the theories contain three different coupled sub-sectors: The structure of the gauge theory sector is similar to the 2d $(0,2)$ GLSMs familiar from the worldsheet formulation of the heterotic string \cite{Witten:1993yc,McOrist:2010ae}. It includes 2d $(0,2)$ chiral and Fermi multiplets charged under the in general abelian and non-abelian gauge group factors originating from
a topologically twisted theory on 7-branes \cite{Schafer-Nameki:2016cfr, Apruzzi:2016iac}.
D3-branes wrapped around curves on the base of the fibration give rise to additional degrees of freedom. These include a particularly fascinating, but largely mysterious sector of Fermi multiplets from the string excitations at the intersection of the D3-branes and the 7-branes \cite{Lawrie:2016axq}.\footnote{The theory on a D3-brane wrapping a curve \cite{Haghighat:2015ega,Lawrie:2016axq} or surface \cite{Martucci:2014ema,Assel:2016wcr} in F-theory is interesting by itself as an example of a gauge theory with varying gauge coupling. An AdS$_3$ gravity dual of an $N=(0,4)$ version has recently been constructed in \cite{Couzens:2017way}.} 
These two sectors are coupled to a 2d $N=(0,2)$ supergravity sector \cite{Lawrie:2016rqe}. 
The construction of 2d $N=(0,2)$ theories has received considerable attention also in other formulations of string theory, most notably via D1 branes probing singularities on Calabi-Yau 4-folds \cite{GarciaCompean:1998kh,Franco:2015tna, Franco:2016nwv,Franco:2016fxm,Franco:2017cjj,Closset:2017yte} and via orientifolds \cite{Forste:1997bd,Font:2004et}.

Various aspects of the non-abelian gauge and the gravitational anomalies in the chiral 2d $(0,2)$ theory obtained via F-theory have already been addressed in \cite{Schafer-Nameki:2016cfr, Apruzzi:2016iac,Apruzzi:2016nfr,Lawrie:2016axq,Lawrie:2016rqe}.
The non-abelian anomalies induced by the chiral fermions in the 7-brane brane gauge sector must be cancelled by the anomalies of the 3-7 modes, as indeed verified in globally consistent examples in \cite{Schafer-Nameki:2016cfr}. The cancellation of all gravitational anomalies for 2d (0,2) supergravities with a trivial gauge theory sector has been proven in \cite{Lawrie:2016rqe} with the help of various index theorems.
Such theories are obtained by F-theory compactification on smooth, generic Weierstrass models. 
On the other hand, the structure of gauge anomalies in the presence of abelian gauge theory factors is considerably more involved, and the subject of this article.

As in higher dimensions, abelian anomalies induced at 1-loop level need not vanish by themselves provided they are consistently cancelled by a two-dimensional version of the Green-Schwarz mechanism. In general 2d $(0,2)$ gauge theories, the structure of the Green-Schwarz mechanism has been laid out in \cite{Adams:2006kb,Quigley:2011pv,Blaszczyk:2011ib} (see \cite{Mohri:1997ef,GarciaCompean:1998kh} for early work). 
In the present situation, the Green-Schwarz mechanism operates at the level of real chiral scalar fields which are obtained by Kaluza-Klein reduction of the self-dual 4-form of Type IIB string theory. They enjoy a pseudo-action which is largely analogous to the pseudo-action of the self-dual 2-tensors in 6d $N=(1,0)$ supergravities and which we parametrise in general terms in section \ref{sec_Anomaliesin2d}. 
As one of our main results we carefully derive this pseudo-action in section \ref{sec_GStermsderivation}, thereby identifying the structure (and correct normalisation) of the anomalous Green-Schwarz couplings. The latter depend on the non-trivial gauge background and imply a classical gauge variance of the right form to cancel the 1-loop abelian gauge anomalies.

A challenge we need to overcome to show anomaly cancellation  is that in absence of a perturbative limit the abelian charges of the 3-7 sector modes are notoriously hard to determine in a microscopic approach. 
Instead of computing the 3-7 anomaly from first principles we extract the anomaly inflow terms onto the worldvolume of the D3-branes in section \ref{sec_GStermsderivation}. To this end we start from the Chern-Simons terms of the 10d effective pseudo-action in the presence of brane sources.
Uplifting this result to F-theory allows us to quantify the contribution of the 3-7 modes in particular to the gauge anomalies and in turn also to deduce the net charge of the 3-7 modes. 

One of our main results is to establish a closed expression for the complete gauge and gravitational anomalies of a 2d $(0,2)$ theory obtained by F-theory compactified on a Calabi-Yau 5-fold.
The resulting conditions for anomaly cancellation are summarized in 
(\ref{gaugeanomalyFtheory}) and (\ref{gravitationalanomalies1}) of section \ref{sec_AnomalyEqu5folds}.
The structure of anomalies reflected in these equations interpolates between 
their analogue in 6d and 4d F-theory vacua:
In 6d F-theory vacua the anomalies are purely dependent on properties of the elliptic fibration, while in 4d they vanish in absence of background flux and depend linearly on the flux background. In 2d F-theory vacua, we find both a purely geometric and a flux dependent contribution to the anomalies. 
 For anomalies to be cancelled, the flux dependent and the flux independent parts of the topological identities  (\ref{gaugeanomalyFtheory}) and (\ref{gravitationalanomalies1}) must in fact hold separately, on any elliptically fibered Calabi-Yau 5-fold and for any gauge background satisfying the consistency relations reviewed in section \ref{sec_FonellCY5}. 
We verify these highly non-trivial anomaly relations in a concrete example fibration for all chirality inducing gauge backgrounds in section \ref{sec_ExampleSU5U1}.

It has already been pointed out that, despite their rather different structure at first sight, the gauge anomalies in 6d and 4d boil down to one universal relation in the cohomology ring of an elliptic fibration over a general base, and similarly for the mixed gauge-gravitational anomalies \cite{Bies:2017abs}.\footnote{By contrast, the purely gravitational anomaly in 6d has no direct counterpart in 4d. See, however, \cite{Grimm:2012yq}.}
This prompts the question if the 2d anomaly relations (\ref{gaugeanomalyFtheory}) and (\ref{gravitationalanomalies1}) are also equivalent to this universal 
relation governing the structure of anomalies in four and six dimensions.
As we will see in section \ref{sec_Comparison6d4d}, assuming the 4d/6d relation of \cite{Bies:2017abs}  implies the flux dependent part of 
(\ref{gaugeanomalyFtheory}) and (\ref{gravitationalanomalies1})
 for a special class of gauge background.
However, it remains for further investigation whether the precise relations extracted  in \cite{Bies:2017abs} on Calabi-Yau 3-folds and 4-folds follow in turn by anomaly cancellation on Calabi-Yau 5-folds in full generality.

\section{Anomalies in 2d $(0,2)$ supergravities} \label{sec_Anomaliesin2d}

Consider an $N=(0,2)$ supersymmetric theory in two dimensions with 
gauge group
\bqa \label{eq:Gsplit}
	G^{\mathrm{tot}}=\prod_{I=1}^{n_G}  G_{I}\times  \prod_{A=1}^{n_{U(1)}} U(1)_A 
\eqa
and matter fields in representations 
\bqa
	\mathbf{R}=(\mathbf{r}^1,\ldots,
	\mathbf{r}^{n_G})_{\underline{q}} \,.
\eqa
Here $\mathbf{r}^I$ denotes an irreducible representation of the simple gauge group factor $G_I$ and $\underline{q}=(q_1,
\ldots ,q_{n_{U(1)}})$ are the charges under the Abelian gauge group factors.
We are interested in the structure of the gauge and gravitational anomalies in such a theory.
These are induced by chiral matter at the 1-loop level. In a general $D$-dimensional quantum field theory, the gauge and gravitational anomalies can be described by a gauge invariant anomaly polynomial of degree $D/2+1$ in the gauge field strength $F$  and the curvature two-form $R$, 
\bqa
\label{poly}
I_{D+2}=\sum_{\mathbf{R},s} n_s(\mathbf{R})I_{s}(\mathbf{R})|_{D+2} \,,
\eqa
where the sum is over all matter fields with spin $s$ which have zero-modes in representation $\mathbf{R}$ with multiplicity  $n_s(\mathbf{R})$.
In particular, a chiral fermion, corresponding to $s=1/2$, contributes with
\bqa
\label{anomaly}
I_{1/2}(\mathbf{R})=- {\rm tr}_\mathbf{R} \, e^{ -F}  \, \hat A({\rm T}) \,,
\eqa 
where $\hat{A}({\rm T})$ is the A-roof genus and $F$ denotes the hermitian gauge field strength. An anti-chiral fermion contributes with the opposite sign. For more details on our conventions we refer to appendix \ref{convention}. 
In $D=2$ dimensions, the 1-loop anomaly polynomial from the charged matter sector is hence a 4-form.
Correspondingly, the anomaly contribution from chiral and anti-chiral fermions in the theory sums up to 
\bqa \label{I4total}
I_{4}=\sum_{\mathbf {R}} (n_+(\mathbf{R}) - n_-(\mathbf{R})  ) \left(-\frac{1}{2}{\rm tr}_{\mathbf R} ( F)^2 + \frac{1}{24} p_1({\rm T}) \,  {\rm dim}({\bf R}) \right) \,,
\eqa
where the first Pontryagin class of the tangent bundle is defined as $p_1({\rm T}) = - \frac{1}{2} {\rm tr} R^2$.
For future purposes we express the anomaly polynomial for the non-abelian, the abelian and the gravitational anomaly as
\bea
I_4|_{G_I} &=&  - {\cal A}_{I}   \,   {\rm tr}_{\bf fund} F_I^2 =   - \frac{1}{2}    \sum_{\mathbf{\mathbf{r}}^I} c^{(2)}_{\mathbf{r}^I}  \,  \chi(\mathbf{r}^I) \,    {\rm tr}_{\bf fund} F_I^2  \label{AI-gen1} \\
I_4|_{ A B} &=&  - {\cal A}_{AB}  \, F^A F^B =  -  \frac{1}{2} \sum_{\mathbf{R}}    q_{A}({\bf R}) \, q_{B}({\bf R})   \,\text{dim}(\mathbf{R})  \, \chi(\mathbf{R})        \,  F^A F^B  \label{AAB-gen1}  \\
I_4|_{\rm grav} &=&    \frac{1}{24}   {\cal A}_{\rm grav}\,    p_1(T)   =   \frac{1}{24} \sum_{\mathbf{R}}\chi(\mathbf{R}) \,  \text{dim}(\mathbf{R})  \,  p_1({\rm T})      \,,
\eea
with $\chi({\bf R})$ denoting the chiral index of zero-modes in representation ${\bf R}$. In the first line we have related the trace in a representation ${\bf r}^I$ of the simple gauge group factor $G_I$ to the trace in the fundamental representation 
via
\bea \label{cr2}
\text{tr}_{\mathbf{r}^I}   F^2=   c^{(2)}_{\mathbf{r}^I} \,    \text{tr}_{\mathbf{fund}}F^2 \,. 
	\eea

In general, the 1-loop induced quantum anomaly need not be vanishing in a consistent theory provided the tree-level action contains gauge variant terms, the Green-Schwarz counter-terms, which cancel the anomaly encoded by $I_{D+2}$. For this cancellation to be possible, 
the 1-loop anomaly polynomial $I_{D+2}$ of the matter sector must factorize suitably.
In two dimensions, the Green-Schwarz counterterms derive from gauge variant interactions of scalar fields. 
The structure of the possible Green-Schwarz terms in a general 2d $N=(0,2)$ supersymmetric field theory has been analyzed in \cite{Adams:2006kb,Quigley:2011pv,Blaszczyk:2011ib} (see \cite{Mohri:1997ef,GarciaCompean:1998kh} for early work).
In this paper, however, we are interested in the specific 2d $N=(0,2)$ effective theory obtained by compactification of F-theory on an elliptically fibered Calabi-Yau 5-fold \cite{Schafer-Nameki:2016cfr,Apruzzi:2016iac}.
In these theories a gauge theory with gauge group (\ref{eq:Gsplit}) is coupled to a 2d $N=(0,2)$ supergravity sector.\footnote{The gauge theory in question arises from spacetime-filling 7-branes. In addition, the compactification contains spacetime-filling D3-branes, but the associated gauge fields are projected out due to $SL(2,\mathbb Z)$ monodromies along the D3-brane worldvolume  \cite{Schafer-Nameki:2016cfr, Lawrie:2016axq}. }
The latter contains a set of 
real axionic scalar fields $c^\alpha$ arising from the Kaluza-Klein (KK) reduction of the F-theory/Type IIB Ramond-Ramond
forms $C_4$ \cite{Lawrie:2016rqe}.\footnote{As discussed in \cite{Lawrie:2016rqe}, these scalars split into $n_+$ chiral and $n_-$ anti-chiral real scalars. Out of these $n_+$ pairs of real chiral and anti-chiral scalars form non-chiral real scalars, which constitute the imaginary part of the bosonic component of a corresponding number of 2d $(0,2)$ chiral multiplets. The remaining $\tau = n_- - n_+$ anti-chiral real scalars form 2d $(0,2)$ tensor multiplets and contribute, together with the gravitino, to the gravitational anomaly at 1-loop level according to the general formulae reviewed in appendix \ref{convention}. This contribution to the 1-loop anomaly is in addition to the classical gauge variance of the Green-Schwarz action discussed in this section.}
 As we will derive in detail in section \ref{sec_GStermsderivation}, their pseudo-action can be parametrized as
\bea \label{2daction1}
S_{\rm GS} = -\frac{1}{4} \int_{\mathbb R^{1,1}} g_{\alpha \beta} \,  H^\alpha \wedge \ast H^\beta - \frac{1}{2} \int_{\mathbb R^{1,1}} \Omega_{\alpha \beta} \,  c^\alpha \wedge X^\beta \,.
\eea
The structure of this action is completely analogous to the well-familiar generalized Green-Schwarz action \cite{Green:1984bx,Sagnotti:1992qw} of self-dual tensor fields in $D = 6$ (see e.g. \cite{Bonetti:2011mw}) and, in fact, $D=10$ dimensions, with the role of the gauge invariant self-dual field strengths being played here 
by the 1-forms $H^\alpha = D c^\alpha$. These are subject to the self-duality condition
\bea \label{selfduality2d}
g_{\alpha \beta} \ast H^\alpha = \Omega_{\alpha \beta} H^\beta \,.
\eea 
The second term in the action constitutes the Green-Schwarz coupling, which is responsible for the non-standard Bianchi identity
\bea
d  H^\alpha = X^\alpha \,,
\eea
where we used that $\Omega_{\alpha \beta}$ is a constant matrix.
The Green-Schwarz couplings will be found to take the form
\bea \label{Xbeta}
X^\beta =  \Theta_A^\beta \, F^A  
\eea
with $F^A$ the field strength associated with the gauge group factor $U(1)_A$ and with $\Theta_A^\beta$ depending on the background flux.
This identifies $H^\alpha$ as
\bea \label{covariant}
H^\alpha = D c^\alpha = d c^\alpha + \Theta_A^\alpha  A^A \,.
\eea
The axionic shift symmetry  of the chiral scalars is gauged by the abelian vector $A^A$ according to the transformation rule
\be
\ba \label{A-gauging}
A^A  &\rightarrow& A^A+d\lambda^A \cr
c^\alpha    &\rightarrow& c^\alpha -\Theta^{\alpha}_{A} \lambda^A
\ea
\ee
such that the covariant derivative $D c^\alpha$
is gauge invariant.
As a result, the pseudo-action picks up a gauge variation of the form
\bea
\delta S_{GS} = \frac{1}{2} \int \Omega_{\alpha \beta} \,   \Theta_A^\alpha \lambda^A \,   X^\beta  =: 2\pi  \, \int_{\mathbb{R}^{1,1}} I_{2}^{(1), {\rm GS}}(\lambda) \,,
\eea
with $I^{(1),{\rm GS}}_{2}$ a gauge invariant 2-form. By the standard descent procedure, it defines an anomaly-polynomial $I^{\rm GS}_4$ encoding the contribution to the total anomaly from the Green-Schwarz sector.
Concretely, the descent equations
\bqa
I^{\rm GS}_4=d I^{\rm GS}_3, \qquad \quad \delta_\lambda I^{\rm GS}_3=d I_{2}^{(1), {\rm GS}}(\lambda) 
\eqa
imply 
\bqa
\label{factor}
2 \pi I^{\rm GS}_4=  \frac{1}{2}  \Omega_{\alpha \beta} X^\alpha X^\beta   = \frac{1}{2}   \Omega_{\alpha \beta}   \Theta_A^\alpha  \Theta_B^\beta  \, F^A   F^B \,.
\eqa
Consistency of the theory then requires that 
\bea
I_4 + I_4^{\rm GS} = 0 \,.
\eea
This is possible only if the non-abelian and gravitational anomalies vanish by themselves and the abelian anomalies factorise suitably.
The resulting constraints on the spectrum take the following  form: 

\begin{subequations} \label{AnomaliesF-theory}
\begin{empheq}[box=\widefbox]{align}
\text{Non-abelian} &: \qquad  
 & \frac{1}{2} \sum_{\mathbf{\mathbf{R}}^I}  \chi(\mathbf{r}^I) \,  c^{(2)}_{\mathbf{r}^I}=0    \label{Non-abelian1}\\
	\text{Abelian}  &:\qquad  
        & \frac{1}{2}\sum_{\bf R} {\rm dim}({\bf R}) \, \chi({\bf R}) \,   q_{A}({\bf R}) \, q_{B}({\bf R})=   \frac1{4\pi}  \Omega_{\alpha \beta}   \Theta_A^\alpha  \Theta_B^\beta  \label{Abelian1}  \\
               \text{Gravitational} &:\qquad 
       &\sum_{\mathbf{R}} \text{dim}(\mathbf{R}) \, \chi(\mathbf{R}) =0\,. \label{Gravity1}
\end{empheq}
\end{subequations}

Note that, unlike in higher dimensions,  the 2d GS mechanism operates entirely at the level of the abelian gauge group factors: 
In $(4k + 2)$ dimensions the analogue of (\ref{covariant}) is the gauge invariant field strength associated with the self-dual rank $(2k+1)$-tensor fields, and the correction term
in the covariant action involves the Chern-Simons $(2k+2)$-forms associated with the gauge and diffeomorphism group. In 2d the Chern-Simons form is proportional to the trace over the gauge connection and must hence be abelian. Therefore the 2d non-abelian and gravitational anomalies from the chiral sector at 1-loop must vanish by themselves; likewise there can be no mixed gravitational-gauge anomalies induced at 1-loop.

Furthermore, let us point out that in the 2d $(0,2)$ theories of the type considered here the gauging (\ref{A-gauging}) of the scalars is directly related to the anomalous Green-Schwarz coupling (\ref{Xbeta}). This is a notable difference to the implementation of the Green-Schwarz mechanism in the more general 2d $(0,2)$ gauge theories of \cite{Adams:2006kb}, where these two are in principle independent.

Before we proceed, we would like to comment on the scalar fields $c^\alpha$. In principle, all of the axionic scalar fields $c^\alpha$ obtained from the Type IIB RR fields $C_p$ can  contribute to the Green-Schwarz mechansim. However, as in 6d and 4d F-theory compactifications, the gauging of the scalar fields from $C_2$ is encoded via a {\rm geometric} St\"uckelberg mechanism in terms of non-harmonic forms, at least in the description via the dual M-theory \cite{Grimm:2011tb}. In this work we will we will only focus on the Green-Schwarz mechanism associated with the scalar fields arising from the RR potential $C_4$, which will be seen to depend on the background flux.

\section{F-theory on elliptically fibered Calabi-Yau five-manifolds} \label{sec_FonellCY5}

In this section we provide some background material on $N=(0,2)$ supersymmetric compactifications of  F-theory to two dimensions. The reader familiar with this type of constructions from \cite{Schafer-Nameki:2016cfr,Apruzzi:2016iac} can safely skip this summary.

\subsection{Gauge symmetries and gauge backgrounds, and 3-branes}

We  consider a 2d $(0,2)$ supersymmetric  theory describing a  vacuum of F-theory compactified on an elliptically fibered Calabi-Yau 5-fold $X_5$ \cite{Schafer-Nameki:2016cfr,Apruzzi:2016iac}  with projection
\be
\pi: X_5\rightarrow B_4 \,.
\ee
The base  $B_4$ is a smooth complex 4-dimensional K\"ahler manifold, which is to be identified with the physical compactification space of F-theory. Via F/M-theory duality, F-theory on $B_4$ is related to the supersymmetric quantum mechanics \cite{Haupt:2008nu} obtained by compactification of M-theory on $X_5$. 

For simplicity we assume that $X_5$ has a global section $[z=0]$ so that it can be described by a Weierstrass equation
\bqa
y^2=x^3+f\, x\, z^4+g\, z^6\,.
\eqa
Here the projective coordinates $[x :  y : z ]$ parametrise the fiber ambient space  $\mathbb{P}_{2,3,1}$ and $f,g$ are sections of the fourth and sixth power of the anti-canonical bundle 
$\bar K$ of the base.
 The discriminant locus 
\bqa
\Delta=4f^3+27g^2=0
\eqa
specifies the location of the 7-branes.
The non-abelian gauge group factors $G_I$ in (\ref{eq:Gsplit}) are associated with 7-branes wrapping divisors $W_I$, which are complex 3-dimensional components of the discriminant locus $\Delta=0$ in the base.
We assume that the Kodaira singularities in the fibre above $W_I$ admit a crepant resolution\footnote{To avoid clutter we will mostly avoid the hat above $\pi$ in the sequel.}
\be
\hat \pi: \hat X_5\rightarrow B_4 \,.
\ee
The resolution replaces the singularity over $W_I$ by a chain of  
 rational curves. After taking into account monodromy effects, which appear for non-simply laced groups, this allows one to identify a collection $\mathbb{P}^1_{i_I}$, $i_I = 1, \ldots, {\rm rk}(\mathfrak{g}_I)$ of independent rational curves in the resolved fiber which can be associated with the simple roots $\alpha_{i_I}$  of the Lie algebra $\mathfrak{g}_I$ underlying $G_I$ in the following sense:
The fibration of $\mathbb P^1_{i_I}$ over  $W_I$  - more precisely of the image of $\mathbb P^1_{i_I}$ under monodromies in the non-simply laced case - defines a resolution divisor $E_{i_I}$ with the property that 
 \bqa
[E_{i_I}]\cdot [\mathbb{P}^1_{j_J}]=-\delta_{IJ} \, C_{i_Ij_J} \,.
\eqa
Here $[E_{i_I}]$ denotes the homology class of the divisor $E_{i_I}$ and unless noted otherwise, all intersection products are taken on $\hat X_5$. The matrix $C_{i_I j_I}$ is the Cartan matrix  of $\mathfrak{g}_I$ (in  conventions where the entries on its diagonal are $+2$).
Via duality with M-theory, M2-branes wrapping the fibral curves  $\mathbb{P}^1_{j_J}$ give rise to states associated with the simple roots $ - \alpha_{i_I}$, and the Cartan $U(1)_{i_I}$ gauge field arises by KK reduction of the M-theory 3-form as 
 \bqa
C_3=A_{i_I}\wedge [E_{i_I}] + ....
\eqa
In this sense the resolution divisors $[E_{i_I}]$ can be identified with the generators ${\cal T}_{i_I}$ of the Cartan subgroup of $G_I$ in the so-called co-root basis, 
whose trace over the fundamental representation of $G_I$ is normalised such that 
\bea
{\rm tr}_{\rm fund} {\cal T}_{i_I} {\cal T}_{j_J} = \delta_{IJ} \,  \lambda_I  \,  \mathfrak{C}_{i_I j_I}  \qquad \quad {\rm with} \qquad \mathfrak{C}_{i_I j_I}=\frac{2}{\lambda_I}  \frac{1}{\langle  \alpha_{j_I}, \alpha_{j_I}  \rangle} \, C_{i_I j_I} \,.
\eea
The quantity $\lambda_I$ denotes the Dynkin index in the fundamental representation and is tabulated in \autoref{Tab_Lambda}. Note that for simply-laced groups $\mathfrak{C}_{i_I j_I} = {C}_{i_I j_I}$.
The geometric manifestation of this identification is the important  relation
\be \label{piEiEj}
\pi_\ast( [E_{i_I}] \cdot [E_{j_J}])=-\delta_{IJ}  \, \mathfrak{C}_{i_I j_I}  \, [W_I]  =   -  {\rm Tr}  \, {\cal T}_{i_I} {\cal T}_{j_J}   \, [W_I]\,,
\ee
where ${\rm Tr}$ is related to the trace in the fundamental representation via
\bea \label{defTr}
{\rm Tr} = \frac{1}{\lambda_I}  \, {\rm tr}_{\bf fund} \,.
\eea
The push-forward  $\pi_\ast( [E_{i_I}] \cdot [E_{j_J}])$ to the base of the fibration is defined by requiring that
\bea
[E_{i_I}]\cdot_{\hat X_5}[E_{j_J}]\cdot_{\hat X_5} [D_\alpha]\cdot_{\hat X_5}[D_\beta] \cdot_{\hat X_5}  [D_\gamma]=\pi_\ast( [E_{i_I}]\cdot_{\hat X_5}[E_{j_J}])\cdot_{B_4} [D^{\rm b}_\alpha]\cdot_{B_4}[D^{\rm b}_\beta] \cdot_{B_4}[D^{\rm b}_\gamma]
\eea
 for any basis of vertical divisors $[D_\alpha] =  \pi^*[D^{\mathrm{b}}_\alpha] $, where $D^{\mathrm{b}}_\alpha$ is a divisor on $B_4$. 
  
 \begin{table}
\begin{center}
\begin{tabular}{|c||c|c|c|c|c|c|c|c|c|}
\hline
$ {\mathfrak g}$ & $A_n$ & $D_n$ & $B_n$ & $C_n$ & $E_6$ & $E_7$ & $E_8$ & $F_4$ & $G_2$ \\ \hline
$ \lambda$         &   $1$   &    $2$  &  $2 $     &  $1$   &   $6$     & $12$   & $60$  & $6$   & $2$   \\ \hline
 \end{tabular}
\caption{Dynkin index of the fundamental representation for the simple Lie algebras. \label{Tab_Lambda}} 
 \end{center} 
 \end{table}

 Each non-Cartan Abelian gauge group factor $U(1)_A$ is associated with a global rational section $S_A$ of $\hat X_5$ in addition to the zero-section $S_0$. To each $S_A$ one can assign an element $[U_A] \in \mathrm{CH}^1(\hat X_5)$ through the Shioda map
\bqa \label{ShiodaUA}
U_A=S_A - S_0-  D_A+  \sum_{i_I}k_{i_I} E_{i_I}  \,.
\eqa
The vertical divisor $D_A$ and the in general fractional coefficients  $k_{i_I}$  are chosen such that $U_A$ satisfies the transversality conditions
\be
\ba
&[U_A]\cdot_{\hat X_5} [D_\alpha]\cdot_{\hat X_5}[D_\beta]\cdot_{\hat X_5}[D_\gamma] \cdot_{\hat X_5} [D_\delta]=0 \qquad [U_A]  \cdot_{\hat X_5}[S_0]  \cdot_{\hat X_5} [D_\alpha]\cdot_{\hat X_5}[D_\beta]  \cdot_{\hat X_5}[D_\gamma]  =0 \cr
&[U_A]  \cdot_{\hat X_5}[E_{i_I}]    \cdot_{\hat X_5} [D_\alpha]\cdot_{\hat X_5}[D_\beta] \cdot_{\hat X_5}[D_\gamma]   =0 \,,
\ea
\ee
which must hold for every vertical divisor $[D_{\alpha}] = \pi^\ast D^{\mathrm b}_{\alpha}$.

In analogy with the relation (\ref{piEiEj}), one can define the so-called height pairing \cite{Park:2011ji,Morrison:2012ei}
 \bea \label{heightpairing}
 \pi_\ast ( [U_A] \cdot_{\hat X_5} [U_B]) =  -    {\rm Tr} \,  {\cal T}_A {\cal T}_B  \,  [D_{AB}]    \, . 
 \eea
The objects  ${\cal T}_A$, $ {\cal T}_B$ are the generators of $U(1)_A$ and $U(1)_B$ and $D_{AB}$ is a divisor on the base of the fibration. 
Unlike the divisor $W_I$, even for $A=B$ this divisor is not one of the irreducible components of the discriminant $\Delta$ (in the sense that $\Delta$ would factorise into the union of various irreducible such $D_{AA}$). Nonetheless, we will see that it plays a very analogous role for the structure of anomalies also for F-theory compactifications to 2d.

A crucial ingredient in F/M-theory compactifications on Calabi-Yau five-folds is the gauge background for the field strength  $G_4 =d C_3$ of the M-theory 3-form potential field.
As in compactifications to four dimensions, the full gauge background is an element of the Deligne cohomology group $H^4_D(\hat X_5, \mathbb Z(2))$ and can be parametrized by equivalence classes of rational complex codimension-2-cycles \cite{Bies:2014sra,Bies:2017fam}, which form the second Chow group $\mathrm{CH}^2(\hat X_5)$. The field strength of $G_4$ as such takes values in 
$H^{4}( \hat{X}_5)$. It is subject to the 
Freed-Witten quantization condition \cite{Witten:1996md}
\be
G_4 + \frac{1}{2} c_2(Y_5) \in H^4(\hat{X}_5,\mathbb Z)\,.
\ee
In order to preserve two supercharges in the M/F-theory compactification on $\hat{X}_5$, the  $(3,1)$ and $(1,3)$ Hodge components of $H^{4}( \hat{X}_5)$ must vanish \cite{Haupt:2008nu} and hence
\be \label{SUSYG422}
G_4 + \frac{1}{2} c_2(Y_5) \in H^4(\hat{X}_5,\mathbb Z) \cap H^{2,2}(\hat{X}_5)\,.
\ee
By F/M-duality, the $G_4$ fluxes are subject to the transversality constraints
\be \label{transverse1}
\int_{\hat X_5} G_4 \wedge S_0 \wedge \pi^\ast\omega_4 = 0 \quad 
{\hbox{and}}
\qquad  \int_{\hat X_5} G_4 \wedge \pi^\ast \omega_6 = 0\,, \qquad \forall  \, \omega_4 \in H^4(B_4), \, \, \omega_6 \in H^6(B_4)\,.
\ee
If this flux satisfies in addition the constraint
\bea \label{gaugeinvariance-a}
\int_{\hat X_5} G_4 \wedge E_{i_I} \wedge \pi^\ast\omega_4 = 0 
\eea
it leaves  the gauge group factor $G_I$ unbroken.

Higher curvature corrections in the M-theory effective action induce a curvature dependent tadpole for the M-theory 3-form $C_3$.
In the dual F-theory these curvature corrections subsume the curvature contributions to the Chern-Simons action of the 7-branes (including, in the perturbative limit, the orientifold planes).
 In a consistent M-theory vacuum this tadpole must be cancelled by the inclusion of background flux $G_4$ and/or by M2-branes wrapping a curve class on $\hat X_5$ determined by the tadpole equation \cite{Haupt:2008nu}. The projection of this curve class to the base $B_4$ describes\footnote{The M2-brane states along the fibral component of this class are related to momentum modes along the circle $S^1$ arising in F/M-theory duality \cite{Lawrie:2016rqe}.} , in the dual F-theory, the class wrapped by background D3-branes filling in addition the extended directions along $\mathbb R^{1,1}$. The projected class is given by \cite{Haupt:2008nu,Schafer-Nameki:2016cfr}
 \bea \label{Cclass}
 [C] =  \frac{1}{24} \pi_\ast c_4(\hat X_5) - \frac{1}{2} \pi_\ast (G_4 \cdot_{\hat X_5} G_4) \,.
 \eea

 \subsection{Matter spectrum from F-theory compactification on CY 5-folds}
 
 The charged chiral matter fields whose contributions to the 1-loop anomalies  we will be studying arise from three sources \cite{Schafer-Nameki:2016cfr,Apruzzi:2016iac}:
 7-brane bulk matter propagating along the non-abelian divisors $W_I$, 7-brane codimension-two matter localised along the intersections of various discriminant components or self-intersections of the discriminant, and finally Fermi multiplets at the pointlike intersection of D3-branes with the 7-branes. Due to the chiral nature of the 2d $(0,2)$ theory, all three types of matter are chiral even for vanishing gauge backgrounds.

The bulk matter fields transform, in the absence of gauge flux, in the adjoint representation of $G_I$. In the dual M-theory quantum mechanics, this matter arises from M2-branes wrapping suitable combinations of resolution $\mathbb P^1_{i_I}$ in the fiber over $W_I$.
For non-vanishing gauge backgrounds, 
which can be described by a non-trivial principal gauge bundle $L$, the original gauge group $G_I$ can be broken into a product of some sub-groups. The spectrum decomposea into irreducible representations $ \mathbf{R}$ of the unbroken gauge factors 
 \bea
 G_I &\rightarrow& H_I  \\
 \mathbf{Adj}(G_I)  &\rightarrow&   \mathbf{Adj}(H_I) \oplus   \bigoplus_{\mathbf{R}}\mathbf{R}
 \eea
Note that if ${\bf R} \neq {\bf \bar  R}$, each representation is accompanied by its complex conjugate. 
The matter fields organise into 2d $(0,2)$ chiral multiplets, which contain one complex boson and a complex chiral Weyl fermion, as well as Fermi multiplets, which contain one complex anti-chiral Weyl fermion. Each of these matter fields is counted by a certain cohomology group on $W_I$ involving the vector bundle $L_{\mathbf{R}}$.
The chiral index of massless matter in a given complex representation, defined as the difference of chiral and anti-chiral fermions in complex representation ${\mathbf{R}}$, is then given by \cite{Schafer-Nameki:2016cfr,Apruzzi:2016iac}
\bea \label{chibulk1a}
\chi({\bf R}) =  - \int_{W_I} c_1(W_I) \left( \frac{1}{12} {\rm rk}(L_{\mathbf R}) \,  c_2(W_I) + \mathrm{ch}_2(L_{\mathbf R}) \right) \,.
\eea
For real representations, this expression is to be multiplied with a factor of $\frac{1}{2}$. In particular, the chiral index of the adjoint representation depends purely on the geometry and takes the form $\chi( \mathbf{Adj}(H_I)) = - \frac{1}{24} \int_{W_I} c_1(W_I)  c_2(W_I) $.

Extra matter states in representation $\mathbf{R}$ of $G^{\rm tot}$    localizes  on complex 2-dimensional surfaces $C_{\mathbf{ R}}$  on $B_4$.
This occurs whenever 
some of the rational curves $\mathbb{P}^1_{i_I}$  in the fiber split over $C_{\mathbf{ R}}$.
Group theoretically, this signifies the splitting of the associated simple roots into weights of representation ${\bf R}$.

The associated charged matter fields arise from M2-branes wrapped on suitable linear combinations of fibral curves over $C_{\mathbf{ R}}$, which in fact span
the weight lattice of the gauge theory.
Hence to each state in representation $\mathbf{ R}$ we can associate a matter 3-cycle $S^a_{\mathbf{ R}}$ which is given by a linear combination of fibral curves over $C_{\mathbf{ R}}$ and carries a weight vector $\beta^a_{i_I}, a=1,...,{\rm dim}(\mathbf{R})$, such that 
 \be
 \pi_\ast ([E_{i_I}]\cdot [S^a_{\mathbf{ R}}])=\beta^a_{i_I}[C_{\mathbf{ R}}] \,.
 \ee
 These matter states also organize both into chiral and Fermi multiplets and are counted by cohomology groups of a vector bundle $L_{\mathbf R}$ which derives from the gauge background.
If the surface $C_{\mathbf{R}}$ on $B_4$ is smooth, the chiral index of this type of matter follows from an index theorem as \cite{Schafer-Nameki:2016cfr,Apruzzi:2016iac}
\be
\ba  \label{chiS2}
\chi({\bf R})
                      &= \int_{C_{\bf R}}  \left(c_1^2(C_{\bf R}) \left(\frac{1}{12} - \frac{1}{8} {\rm rk}(L_{\bf R}) \right) + \frac{1}{12} c_2(C_{\bf R}) + \left(\frac{1}{2} c_1^2(L_{\bf R}) - c_2(L_{\bf R})\right) \right).       
\ea\ee 
Otherwise one has to perform a suitable normalisation in order to be able to apply the index theorem, and this will lead to correction terms as exemplified in \cite{Schafer-Nameki:2016cfr}.

 The third type of massless matter arises from 3-7 string states at the intersection of the 7-branes with the spacetime-filling D3-branes wrapping the curve class $[C]$ in (\ref{Cclass}).
 Matter in the 3-7 sector comes in 2d (0,2) Fermi multiplets \cite{Schafer-Nameki:2016cfr,Apruzzi:2016iac}. 
 In purely perturbative setups, each intersection point of $[C]$ with one of the D7-branes carries a single Fermi multiplet in the fundamental representation of the D7-brane gauge group.
 However, monodromy effects along the 3-brane worldvolume considerably obscure such  a simple interpretation of the 3-7 modes in non-perturbative setups \cite{Schafer-Nameki:2016cfr, Lawrie:2016axq}. As one of our results, we will see how the structure of 2d anomalies sheds new light on the structure of 3-7 modes, including, in particular, their charges under the non-Cartan abelian gauge factors.

\section{Anomaly equations in F-theory on Calabi-Yau 5-folds } \label{sec_AnomalyEqu5folds}

In this section we present closed expressions for the anomaly cancellation conditions in 2d $(0,2)$ F-theory vacua. We begin in section \ref{sec_gaugean} by deriving a formula for the chiral index of charged matter states in the presence of 4-form flux $G_4$ in the dual M-theory, which is uniformly valid for the bulk and the localised 7-7 modes. 
We also shed some more light on the counting of 3-7 modes. Together with the Green-Schwarz counterterms this leads to formula (\ref{gaugeanomalyFtheory}) for the cancellation of all gauge anomalies.
In section \ref{gravitationalanomaly}   we extend the gravitational anomaly cancellation conditions of  \cite{Lawrie:2016rqe} to situations with non-trivial 7-branes and fluxes, leading us to condition (\ref{gravanomA-general}).

\subsection{Gauge anomalies, Green-Schwarz terms and the 3-7 sector} \label{sec_gaugean}

Recall from the previous section that in this paper we assume the existence of a smooth crepant resolution $\hat X_5$, which describes the dual M-theory on its Coulomb branch. This forces us, as usual in this context, to restrict ourselves to Abelian gauge backgrounds $G_4$. In particular, the vector bundles appearing in the expressions (\ref{chibulk1a})
and (\ref{chiS2}) are complex line bundles. 

For simplicity of presentation we first assume that the gauge flux $G_4$ does not break any of the non-abelian gauge group factors.
The chiral index (\ref{chiS2}) of the localised matter can be split into a purely geometric and a flux dependent contribution
\bea
\chi({\bf R})   &=&   \chi_{\rm geom}({\bf R})   + \chi_{\rm flux}({\bf R}) \nonumber \\
\chi_{\rm geom}({\bf R}) &=&  - \frac{1}{12} \int_{C_{\bf R}} \mathrm{ch}_2(C_{\bf R})  = \frac{1}{12}  \int_{C_{\bf R}}  c_2(C_{\bf R})  - \frac{1}{2} c_1^2(C_{\bf R} )  \label{chigeom1}\\
\chi_{\rm flux}({\bf R})     &=& \int_{C_{\bf R}} \frac{1}{2} c_1^2(L_{\bf R}) \,. \nonumber 
\eea
We stress that this expression is correct provided the matter 2-cycles $C_{\bf R}$ on $B_4$ are smooth. The line bundle $L_{\bf R}$ on $C_{\bf R}$ to which a state with weight vector $\beta^a({\bf R}))$ couples is obtained from $G_4$ by first integrating $G_4$ over the fiber of the matter 3-cycle $S^a_\mathbf{R}$ and then projecting onto the surface $C_{\bf R}$. This gives rise to a divisor class on $C_{\mathbf{R}}$ which is to be identified, similarly to the procedure in F-theory on Calabi-Yau 4-folds \cite{Bies:2014sra,Bies:2017fam}, with
\bea \label{c1LR}
c_1(L_{\bf R}) = \pi_\ast (G_4 \cdot S^a_{\bf R}) \,.
\eea
Note that for gauge invariant flux, the result is the same for each of the matter 3-cycles $S^a_{\bf R}$ and hence correctly defines the line bundle associated with representation ${\mathbf{R}}$.
This allows us to rewrite $\chi_{\rm flux}({\bf R})  $ explicitly in terms of $G_4$ as
\bea
\chi_{\rm flux}({\bf R})  = \frac{1}{2}  \pi_\ast (G_4 \cdot S^a_{\bf R})  \cdot_{C_{\bf R}} \pi_\ast (G_4 \cdot S^a_{\bf R}) \,,
\eea
where $\cdot_{C_{\bf R}}$ denotes the intersection product on $C_{\bf R}$.

Next, consider the bulk modes.
For gauge invariant flux, this sector contributes only states in the adjoint representation of $G_I$ (which due to the quadratic nature of the anomalies nonetheless contribute to the anomaly), and according to (\ref{chibulk1a}) their chiral index is given by
\bea \label{chibadj1}
\chi_{\rm bulk}({\bf R} = {{\rm \bf adj}_I}) = - \frac{1}{24} \int_{W_I} c_1(W_I)  c_2(W_I) \,.
\eea
It is useful to note that $\chi_{\rm bulk}({\bf R})$ is formally identical to the flux-independent part of the chirality of a localised state whose matter locus is given by the canonical divisor  on $W_I$, i.e. the complex 2-cycle on $W_I$ in the class
\bea
[C_{\rm can}] = -c_1(W_I) = + c_1(K_{W_I}) \,. \,
\eea
Indeed, by adjunction, using the short exact sequence
\bea
0 \rightarrow T_{C_{\rm can}} \rightarrow   T_{W_I} \rightarrow N_{C_{\rm can}/W_I} \rightarrow 0
\eea 
and the resulting relation
\bea
c(T_{C_{\rm can}}) = c( T_{W_I}) / c(N_{C_{\rm can}/W_I}) =  (1 + c_1(W_I) + c_2(W_I)) / (1 -c _1(W_I) ) ,
\eea
one computes 
\bea
c_1(C_{\rm can}) &=& 2 c_1(W_I) \\
c_2(C_{\rm can}) &=& c_2(W_I)  + 2 c_1^2(W_I) \,.
\eea
This implies that 
\bea
\int_{C_{\rm can}} \frac{1}{12} (c_2(C_{\rm can})  - \frac{1}{2} c_1^2(C_{\rm can}) ) = -\frac{1}{12}  \int_{W_I} c_1(W_I) \cdot  c_2(W_I).
\eea
 The additional factor of $\frac{1}{2}$ in (\ref{chibadj1}) is due to the fact that the adjoint is a real representation. 
More generally, and  in complete analogy to the description of bulk modes in compactifications on Calabi-Yau 4-folds \cite{Bies:2017fam}, we can associate to a bulk matter state associated with the root $\rho_I$ the 3-cycle
\bea \label{Srho}
S^{{\rho}_I} = \sum_{i_I} \hat a_{i_I}  E_{i_I} |_{K_{W_I}} \,.
\eea  
The parameters $\hat a_{i_I}$ are related to the coefficients in the expansion of the root $\rho_I$ in terms of the simple roots $\alpha_{i_I}$.\footnote{For simply laced Lie algebras, $\rho_I = \sum_{i_I} \hat a_{i_I} \alpha_{i_I}$. For non-simply laced Lie algebras, fractional corrections must be included to take into account monodromy effects, as explained e.g. in appendix A of \cite{Park:2011ji}.}
Geometrically, the fiber of $S^{{\rho}_I} $ is given by the corresponding linear combination of fibral rational curves $\mathbb P^1_{i_I}$. An M2-brane wrapped along this linear combination of fibral curves gives rise to a state whose Cartan charges are given precisely by the root  $\rho_I$.
For gauge invariant flux satisfying (\ref{gaugeinvariance-a}), the line bundle $\pi_\ast (S^{{\rho}_I} \cdot G_4)$ vanishes by construction. Hence the expression for the bulk and the localised chirality are completely analogous and both types of matter will from now on be treated on the same footing.

This conclusion persists if the gauge background breaks some or all of the simple gauge group factors $G_I$. In this case, the adjoint representation for the bulk matter or the representations  associated with the localised matter decompose into irreducible representations of the unbroken subgroup. 
The operation (\ref{c1LR}) now leads to a well-defined line bundle for each of these individual representations, for bulk and localised matter alike.

Next, we consider the contribution from the 3-7 modes. As it turns out, to each representation ${\bf R}$ one can associate a divisor $D_{37}({\bf R})$ on $B_4$ such that the chiral index of 3-7 states in representation  ${\bf R}$ is given by
\bea \label{D37notation}
\chi_{3-7}({\bf R}) = -  \Big( \frac{1}{24} \pi_\ast (c_4(\hat X_5)) - \frac{1}{2} \pi_\ast(G_4 \cdot G_4) \Big) \cdot_{B_4} D_{37}({\bf R}) \,.
\eea
The expression in brackets is the curve class $[C]$, defined in (\ref{Cclass}), wrapped by the spacetime-filling D3-branes.
For instance, for a perturbative gauge group $G_I = SU(N)$, each intersection point of $[C]$ with the 7-brane divisor $W_I$ hosts a (negative chirality) Fermi multiplet in representation
${\bf R} = ({\bf N})$ \cite{Schafer-Nameki:2016cfr,Apruzzi:2016iac}  and  therefore $D_{37}({\bf R} = ({\bf N})) = W_I$. 
For non-perturbative gauge groups and for Abelian non-Cartan groups $U(1)_A$ determining the representation and charge of the 3-7 strings from first principles is more obscure due to  subtle $SL(2,\mathbb Z)$ monodromy effects on the worldvolume of the D3-brane along $C$ \cite{Lawrie:2016axq}.
However, in the next section we will derive that in the presence of extra $U(1)_A$ gauge group factors  the net contribution to  the $U(1)_A - U(1)_B$  anomaly (\ref{AAB-gen1})
from the 3-7 sector takes the form 
\bea
{\cal A}_{AB}|_{3-7}   &=&    \frac{1}{2} \sum_{{\bf R}, 3-7}     q_A({\bf R}) \,   q_B({\bf R})  \, {\rm dim}({\bf R}) \,   \chi_{3-7}({\bf R})  \\
&=&    \frac{1}{2} \, \Big( \frac{1}{24} \pi_\ast (c_4(\hat X_5)]) - \frac{1}{2} \pi_\ast(G_4 \cdot G_4) \Big) \cdot_{B_4} \pi_{\ast} (U_A \cdot U_B) \,.  \label{37ABanomaly}
\eea
Here we recall that $U_A$ and $U_B$ generate the respective $U(1)$ factors via the Shioda map (\ref{ShiodaUA}) and that 
the height-pairing $\pi_{\ast} (U_A \cdot U_B)$ had been introduced in (\ref{heightpairing}). 
More generally, our results imply that the right-hand side correctly captures the contribution to the anomaly also of the Cartan $U(1)$ group for non-perturbative gauge groups.  
Let us introduce the notation
 \bea
{\rm span}_{\mathbb C} \{ {\mathfrak F}_{\Sigma} \} =  {\rm span}_{\mathbb C} \{E_{i_I}, U_A\}
\eea
to collectively denote set of divisors generating any of the Cartan $U(1)_{i_I}$ or non-Cartan $U(1)_A$ gauge symmetries.
Then our claim is that the contribution to the gauge anomaly due to 3-7 modes can be summarized as
\bea
{\cal A}_{\Lambda \Sigma}|_{3-7} =   \frac{1}{2} \sum_{{\bf R}, a}\beta^a_\Lambda({\bf R})  \, \beta^a_\Sigma({\bf R}) \, \chi_{3-7}({\bf R}) =    \frac{1}{2} \, \Big( \frac{1}{24} \pi_\ast (c_4(\hat X_5)]) - \frac{1}{2} \pi_\ast(G_4 \cdot G_4) \Big) \cdot_{B_4} \pi_\ast (\mathfrak{F}_\Lambda \cdot \mathfrak{F}_\Sigma)  \,.\label{37ABanomaly-b}
\eea
If the index $\Lambda = i_I$ refers to a Cartan $U(1)_{i_I}$,  the object $\beta^a_{i_I}({\bf R})$ denotes the weights  associated with representation ${\bf R}$ with respect this $U(1)_{i_I}$, and for $\Lambda = A$ we define $\beta^a_{A}({\bf R}) = q_A({\bf R})$.
We will come back to the interpretation of this formula at the end of this section.

As the final ingredient we will derive, in section \ref{sec_GStermsderivation}, the Green-Schwarz counterterms  appearing on the righthand side  of (\ref{Abelian1}).
These are found to be purely flux dependent and of the form
 \bea \label{GSclaim}
\frac1{4\pi}  \Omega_{\alpha \beta}   \Theta_\Sigma^\alpha  \Theta_\Lambda^\beta   =  \frac{1}{2} \pi_\ast (G_4 \cdot \mathfrak{F}_\Sigma) \cdot_{B_4}   \pi_\ast (G_4 \cdot \mathfrak{F}_\Lambda ) \,.
\eea
For instance, if we let $\mathfrak{F}_\Lambda = U_A$, $\mathfrak{F}_\Sigma = U_B$ refer to non-Cartan Abelian groups, then this describes the Green-Schwarz counterterms for the $U(1)_A- U(1)_B$ anomalies.
For $\mathfrak{F}_\Lambda = E_{i_I}$, $\mathfrak{F}_\Sigma = E_{j_I}$, the right-hand side is non-vanishing only if the gauge background $G_4$ breaks the simple gauge group factors $G_I$ and $G_J$, in which case it computes the counterterms for the $U(1)_{i_I} - U(1)_{j_J}$ anomaly. For gauge invariant flux, on the other hand, no such Green-Schwarz terms are induced, in agreement with expectations.

With this preparation we can now rewrite the gauge anomaly equations (\ref{Non-abelian1}), (\ref{Abelian1}) in a rather suggestive form.
Since the anomaly equations must hold for arbitrary gauge background $G_4$ 
and since the flux independent terms only give a constant off-set, the flux dependent and the flux independent contributions to the anomalies must vanish separately.  
The requirement (\ref{Non-abelian1}), (\ref{Abelian1})  of cancellation of all gauge anomalies therefore  results in two independent identities:

\begin{subequations} \label{gaugeanomalyFtheory}
\begin{empheq}[box=\widefbox]{align}
0 &=\sum_{{\bf R},a}   \beta^a_\Lambda({\bf R}) \, \beta^a_\Sigma({\bf R})    \int_{C_{\bf R}} \mathrm{ch}_2(C_{\bf R})    - \frac{1}{2}  \,  \pi_\ast (c_4(\hat X_5)) \cdot  \pi_\ast (\mathfrak{F}_\Lambda \cdot \mathfrak{F}_\Sigma)     \label{gauge-geom1} \\
0 &=  \sum_{{\bf R},a}   \beta^a_\Lambda({\bf R}) \, \beta^a_\Sigma({\bf R})  \,  \pi_\ast (G_4 \cdot S^a_{\bf R})  \cdot_{C_{\bf R}} \pi_\ast (G_4 \cdot S^a_{\bf R})  \nonumber  \\
& - \big( \pi_\ast (G_4 \cdot G_4) \cdot_{B_4}   \pi_\ast (\mathfrak{F}_\Lambda \cdot \mathfrak{F}_\Sigma)  \label{gauge-flux1} \\
& \phantom{-(}+     \pi_\ast (G_4 \cdot \mathfrak{F}_\Sigma) \cdot_{B_4}   \pi_\ast (G_4 \cdot \mathfrak{F}_\Lambda ) {+  \pi_\ast (G_4 \cdot \mathfrak{F}_\Lambda) \cdot_{B_4}   \pi_\ast (\mathfrak{F}_\Sigma \cdot G_4) \big) \nonumber \,.} 
\end{empheq}
\end{subequations}
The two terms in (\ref{gauge-geom1}) respectively represent the flux independent anomaly contribution from the 7-7 sector, (\ref{chigeom1}), and from the 3-7 sector, (\ref{37ABanomaly-b}).
In (\ref{gauge-flux1}) we have collected the flux dependent 3-7 and the Green-Schwarz contribution to the anomaly  in the brackets in the second and third line to illustrate the striking formal similarity between them. We will understand this similarity in the next section.

Let us now come back to the interpretation of (\ref{37ABanomaly-b}).
For $\mathfrak{F}_\Lambda = E_{i_I}$, $\mathfrak{F}_\Sigma = E_{j_I}$ this equation allows us to deduce the net contribution to the anomalies due to 3-7 strings charged under the non-abelian gauge group factors, which, as noted already, can be rather obscure due to monodromy effects.
To interpret this expression, recall the crucial identity (\ref{piEiEj}).
If we assume that each geometric  intersection point $[C] \cdot_{B_4} W_I$ hosts an (anti-chiral) Fermi multiplet in representation ${\bf R}$, then  for consistency this representation must satisfy
\bea \label{Cijbeta}
\sum_a \beta^a_{i_I}({\bf R} ) \beta^a_{j_I}({\bf R} )  \stackrel{!}{=}    \mathfrak{C}_{i_I j_I}  \,.
\eea
This is to be contrasted with the fact that for any representation ${\bf R}$ of a simple group $G_I$
\bea
\sum_a \beta^a_{i_I}({\bf R} ) \beta^a_{j_I}({\bf R} )   = {\rm tr}_{\bf R} {\cal T}_{i_I} {\cal T}_{j_I} = \lambda_I \,  c_{\mathbf{R}}^{(2)} \,  \mathfrak{C}_{i_I j_I}  
\eea
with ${\cal T}_{i_I}$ denoting the Cartan generators in the coroot basis.
The Dynkin index  $\lambda_I$  for the fundamental representation of $G_I$ is collected, for all simple groups, in \autoref{Tab_Lambda}, and $c_{\mathbf{R}}^{(2)}$ normalizes the trace with respect to the fundamental representation as in (\ref{cr2}).
By definition, the smallest value of $c_{\mathbf{R}}^{(2)}$ occurs for the fundamental representation $c_{\mathrm{\bf fund}}^{(2)} =1$. Hence unless $\lambda_I = 1$ or $\lambda_I = 2$, the interpretation in terms of 3-7 modes necessarily involves 'fractional' Fermi multiplets.\footnote{The case $\lambda_I = 2$ requires, for consistency, that the  fundamental representation be real and hence contributes with a factor of $\frac{1}{2}$ to compensate for  $\lambda_I$.  \autoref{Tab_Lambda} confirms that this is indeed the case for all simple algebras with $\lambda_I = 2$.} This is in agreement with the observation of \cite{Schafer-Nameki:2016cfr} that e.g. for $G_I = E_6$, the net contribution to the anomaly
from the 3-7 sectors corresponds to that of a $\frac{1}{6}$-fractional Fermi multiplet per intersection point.

\subsection{Gravitational Anomaly}
 \label{gravitationalanomaly}

The gravitational anomaly for F-theory compactified on a smooth Weierstrass model $X_5$ without any 7-brane gauge group and background flux has already been discussed in \cite{Lawrie:2016rqe}.
The anomaly polynomial receives contributions from the moduli sector, from the 2d $(0,2)$ supergravity multiplet as well as from the 3-7 sector,
\bea
I_{4, \rm grav} &=&  \frac{1}{24} p_1(T) \left( {\cal A}_{\rm grav}|_{\rm mod} + {\cal A}_{\rm grav}|_{\rm uni} +  {\cal A}_{\rm grav}|_{3-7}  \right) \\
{\cal A}_{\rm grav}|_{\rm mod} &=& -\tau(B_4) + \chi_1(X_5) - 2 \chi_1(B_4), \label{Amod} \\ 
{\cal A}_{\rm grav}|_{\rm uni} &=& 24 \\
 {\cal A}_{\rm grav}|_{3-7}& =& - 6 c_1(B_4) \cdot [C] \,.
\eea
Note that ${\cal A}_{\rm grav}|_{\rm mod}$ includes what would be called in Type IIB language the contributions from the closed string moduli sector, from the moduli associated with the 7-branes (which however by assumption carry no gauge group), and from $\tau(B_4)$ many 2d $(0,2)$ tensor multiplets. 
Here 
\bea
\chi_q(M) = \sum_{p=1}^{{\rm dim}(M)} (-1)^p h^{p,q}(M) 
\eea
and 
\be
\chi_1(X_5)=-\frac1{24}\int_{X_5} c_5(X_5)=\int_{B_4}(90c_1^4+3c_1^2c_2-\frac12c_1c_3)
\ee 
with $c_i = c_1(B_4)$.
Furthermore the signature $\tau(B_4)$ counts the difference of self-dual and anti-self-dual 4-forms on $B_4$ and is related to the Hodge numbers of $B_4$ as 
\bea
\tau(B_4) = b_4^+(B_4) -  b_4^-(B_4) = 48 + 2 h^{1,1}(B_4) + 2 h^{3,1}(B_4) - 2 h^{2,1}(B_4) \,.
\eea
 The D3-brane class appearing fixed by the tadpole on a smooth Weierstrass model without flux is  $[C]=\frac1{24} \pi_\ast c_4(X_5)$.
As shown in  \cite{Lawrie:2016rqe} with the help of various index theorems, the total anomaly can be evaluated as
\bea \label{I4gravWeiersmooth}
I_{4, \rm grav} = \frac{1}{24} p_1(T)  \left(  -24  \chi_0(B_4) + 24 \right) \equiv 0 \,,
\eea
where the last equality holds because $h^{0,i}(B_4)$ for $i \neq 0$ if $B_4$ is to admit a smooth Calabi-Yau Weierstrass fibration over it.

Suppose now that the fibration contains in addition a non-trivial 7-brane gauge group and charged 7-7 matter, and let us also switch a non-trivial flux background $G_4$.
For simplicity assume first that the supersymmetry condition that $G_4$ be of pure $(2,2)$ Hodge type \cite{Haupt:2008nu} does not constrain the moduli of the compactification. 
In analogy with $G_4$ flux on Calabi-Yau 4-folds, this is guaranteed whenever 
$G_4 \in H^{2,2}_{\rm vert}(\hat X_5)$, the primary vertical subspace of  $H^{2,2}(\hat X_5)$ generated by products of $(1,1)$ forms.\footnote{The space of $(2,2)$ forms on Calabi-Yau 5-folds deserves further study beyond the scope of this article. In particular it remains to investigate in more detail whether a similar split into horizontal and vertical subspaces exists as on Calabi-Yau 4-folds. In any event if $G_4$ is a sum of $(2,2)$ forms obtained as the product of two $(1,1)$ forms, the Hodge type does not vary.}
In this situation the gravitational anomaly generalizes as follows:
First, we must now work on the resolution $\hat X_5$ of the singular Weierstrass model describing the more general 7-brane configuration. 
In particular the D3-brane curve class changes to $[C] = \frac{1}{24} \pi_\ast c_4(\hat X_5) - \frac{1}{2} \pi_\ast (G_4 \cdot G_4)$ with $c_4(\hat X_5)$ evaluated on the resolved space $\hat X_5$.
Second, we must add the anomaly contribution from the non-trivial 7-7 sector. 
This sector includes the localised matter in some representation ${\bf R}$  of the total gauge group as well as the bulk matter in the adjoint representation (or its decomposition if the flux background breaks the non-abelian gauge symmetry). Each massless multiplet in the bulk sector contributes ${\rm dim}(\bf adj)$ many states to the anomaly. Of these, ${\rm rk}(G)$ many states are associated with the Cartan subgroup of the gauge group and are in fact encoded already in the contribution from the 'moduli sector'. More precisely, if we replace in (\ref{Amod}) the contribution $\chi_1(X_5)$ by $\chi_1(\hat X_5)$, the resulting expression ${\cal A}_{\rm grav}|_{\rm mod}$ now includes the anomaly from the ${\rm rk}(G) = h^{1,1}({\hat X_5}) - (h^{1,1}(B_4) -1)$
many vector multiplets associated with the Cartan subgroup as well as the 'open string moduli' in the Cartan, which enter the values of $h^{1,p}(\hat X_5)$.
As a result, the total gravitational anomaly polynomial is now
\be
I_{4, \rm grav} =  \frac{1}{24} p_1(T) \left(  {\cal A}_{\rm grav}|_{\rm 7-7} + {\cal A}_{\rm grav}|_{\rm mod} + {\cal A}_{\rm grav}|_{\rm uni} +  {\cal A}_{\rm grav}|_{3-7}  \right) 
\ee
with the individual contributions
\bea
{\cal A}_{\rm grav}|_{7-7} &=& \sum_{{\bf R}} {\rm dim}({\bf  R}) \chi({\bf R})     -   {\rm rk}(G) \chi({\bf adj}) \\
{\cal A}_{\rm grav}|_{\rm mod} &=& -\tau(B_4) + \chi_1(\hat X_5) - 2 \chi_1(B_4), \\ 
{\cal A}_{\rm grav}|_{\rm uni} &=& 24 \\
 {\cal A}_{\rm grav}|_{3-7}& =& - 6 c_1(B_4) \cdot \left(\frac{1}{24} \pi_\ast (c_4(\hat X_5)) - \frac{1}{2} \pi_\ast (G_4 \cdot G_4) \right)\,.
\eea
Note that  the topological invariants  $\chi_1(\hat X_5)$ and $ c_4(\hat X_5)$
contain correction terms in addition to the base classes appearing for the case of a smooth Weierstrass model which depend on the resolution divisors and extra sections (if present).

The vanishing of the total gravitational anomaly implies that these individual contributions must cancel each other,
\bea \label{gravanomA-general}
 {\cal A}_{\rm grav}|_{\rm 7-7} + {\cal A}_{\rm grav}|_{\rm mod} + {\cal A}_{\rm grav}|_{\rm uni} +  {\cal A}_{\rm grav}|_{3-7}  = 0 \,.
\eea
This leads to a set of topological identities which must hold for every resolution $\hat X_5$ of an elliptically fibered Calabi-Yau 5-fold, and for every consistent configuration of background fluxes thereon, as specified above.
Note that the flux background enters not only through the 3-brane class in  ${\cal A}_{3-7}$, but also because the chiral indices in the 7-brane sector split as $\chi({\bf R}) = \chi({\bf R})|_{\rm geom} + \chi({\bf R})|_{\rm flux}$ as in (\ref{chigeom1}). 
In principle, if the Hodge type of $G_4$ were to vary over the moduli space, the supersymmetry condition $G_4 \in H^{2,2}(\hat X_5)$ would induce a potential for some of the moduli \cite{Haupt:2008nu} and hence modify the number of uncharged massless fields. According to our assumptions, this does not occur for the choice of flux considered here and the uncharged sector contributes to the anomaly as above.

 Then the anomaly equations split into the independent sets of equations
\begin{subequations} \label{gravitationalanomalies1}
\begin{empheq}[box=\widefbox]{align}
&& \sum_{{\bf R}} {\rm dim}({\bf  R}) \chi({\bf R}) |_{\rm geom}    -   {\rm rk}(G) \chi({\bf adj}) |_{\rm geom}  -\tau(B_4) + \chi_1(\hat X_5) - 2 \chi_1(B_4) + 24 \nonumber  \\
&& - \frac{1}{4} c_1(B_4) \cdot \left( \pi_\ast c_4(\hat X_5)\right) = 0 \label{gravgeom1} \\
&& -6c_1\cdot \pi_\ast(G_4\cdot G_4)=  \sum_{\mathbf{R},a}    \pi_\ast(G_4\cdot S^a_\mathbf{R})\cdot_{C_{\mathbf{R}}}\pi_\ast(G_4\cdot S^a_\mathbf{R}) \label{gravflux1}
\end{empheq}
\end{subequations} 
 
 In the second equation, which accounts for the flux dependent anomaly contribution, we do not need to treat the 7-brane states in the Cartan separately as their chirality is not affected by the flux background.
 
The flux independent contribution can be analysed further if the fibration $\hat X_5$ is smoothly connected to a smooth Weierstrass model $X_5$. In the terminology of \cite{Morrison:2014lca}, this means that the F-theory model does not contain any non-Higgsable clusters and hence after the blowdown of the resolution divisors the gauge symmetry can be completely Higgsed.
In that case we know already from (\ref{I4gravWeiersmooth}) that the anomalies on the resulting smooth Weierstrass model $X_5$ cancel for $G_4 =0$.   
Let us therefore define
\bea
\label{DeltaC}
\Delta[C] &=&    \frac 1{24} \left(\pi_\ast c_4(\hat X_5) -  \pi_\ast c_4(X_5) \right) =   \frac 1{24} c_4(\hat X_5)|_{B_4}-(15c_1^3+\frac12c_1c_2)  \\
\Delta\chi_1 &=& -\frac{1}{24} \left( \pi_\ast c_5(\hat X_5)  -  \pi_\ast c_5(X_5) \right)    =  -\frac{1}{24} \pi_\ast c_5(\hat X_5)-(90c_1^4+3c_1^2c_2-\frac12c_1c_3) \,.
\eea
The anomaly equations can then be rewritten as 
\begin{subequations} \label{gravitationalanomalies}
\begin{empheq}[box=\widefbox]{align}
-6c_1\cdot \Delta[C]  +\Delta\chi_1 =& -\frac1{12}\sum_{\mathbf{R}}\text{dim}(\mathbf{R})\int_{C_{\mathbf{R}}} {\rm ch}_2(C_{\mathbf{R}}) \label{curvgravanomalies}\nonumber  \\
& + \frac{1}{12} {\rm rk}(G) \int_{C({\bf adj})} {\rm ch}_2(C({\bf adj})) \\
 -6 c_1\cdot \pi_\ast(G_4\cdot G_4)=&   \sum_{\mathbf{R},a} \pi_\ast(G_4\cdot S^a_\mathbf{R})\cdot_{C_{\mathbf{R}}}\pi_\ast(G_4\cdot S^a_\mathbf{R}) \label{gravitationalanomalies2}
\end{empheq}
\end{subequations} 

It is interesting to speculate about the effect of $G_4$ fluxes which are not automatically of $(2,2)$ Hodge type. 
The supersymmetry condition (\ref{SUSYG422}) is reflected in a dynamical potential which is expected to render some of the supergravity moduli massive \cite{Haupt:2008nu}.
The resulting change in the gravitational anomaly compared to the fluxless geometry must be compensated by a suitable modification of the remaining uncharged spectrum.
Indeed, the flux contributes at the same time to the D3-brane tadpole and hence changes the D3-brane curve class $[C]$ compared to the fluxless compactification.
This changes the number of massless Fermi multiplets in the 3-7 sector.  
The net number of moduli stabilized in the presence of flux must equal the change in the number of 3-7 modes. This interesting effect has no analogue in 6d or 4d F-theory vacua: In 6d there is no background flux, and in 4d there is no purely gravitational anomaly.

\section{Derivation of the Green-Schwarz terms and 3-7 anomaly} \label{sec_GStermsderivation}

In this section we derive the two key results of this paper, the form and correct overall normalization of the 2d Green-Schwarz terms and the contribution to the gauge anomalies from the 3-7 string sector.
As we will see, both can be obtained in a very compact manner directly from the gauging of the Type IIB Ramond-Ramond 4-form in the presence of source terms.
The gauging of the Ramond-Ramond forms in the presence of brane sources is standard \cite{Green:1996dd,Cheung:1997az,Minasian:1997mm}, and a similar ten-dimensional approach to determining the gauging in a compactification has been taken in \cite{Martucci:2015oaa,Martucci:2015dxa}.
We will first derive this gauging in an orientifold limit and describe its implications for the Green-Schwarz terms and its relation to the 3-7 anomalies. We then uplift the result to F-theory on an elliptically fibered Calabi-Yau, which is valid beyond the perturbative limit. We close this section by making contact with the 2d effective action laid out in section \ref{sec_Anomaliesin2d}.

\subsection{10d Chern-Simons terms}

Consider a Type IIB orientifold compactification on  a Calabi-Yau 4-fold $X_4$, with spacetime-filling D7-branes and O7-planes associated with a holomorphic orientifold involution  $\sigma: X_4\rightarrow X_4$. To simplify the presentation we omit orientifold invariant D7-branes and only consider  D7-branes as  pairs D7$_a$, D7$_{a'}$ wrapping effective divisors $D_a$ and $D_{a'} = \sigma_{\ast}(D_a)\neq D_a$ on $X_4$. The  cohomology class Poincar\'e dual to $D_a$ will be denoted by $[D_a]$. The field strength on the D$7_a$-brane is denoted as ${\mathbf F}_a$ with ${\mathbf F}_{a'}=-\sigma_{\ast} ({\mathbf F}_a)$. In addition we allow for spacetime-filling D3-branes and their image wrapping 
curves $C_i$ and $C_{i'}$ on $X_4$. Our conventions for the effective action of the supergravity fields and the branes are summarized in appendix \ref{app_conventions}. 
The 10d supergravity action in the presence of 7-branes and 3-branes and after taking the orientifold quotient takes the form\footnote{The overall factor of $\frac{1}{2}$ results from the orientifold quotient.}
\be
\label{action} 
S=\frac 12 \left(    S_{\rm IIB}+ \sum_{a}  (S_a^{\rm D7}+S_{a'}^{\rm D7})  +  S^{\rm O7} + \sum_{i}  (S_i^{\rm D3}+S_{i'}^{\rm D3})    \right) \,.
\ee
We are interested in the gauging  of the RR 4-form potential $C_4$. Prior to taking into account the source terms due to the branes, its associated field strength is\footnote{Strictly speaking, the $SL(2,\mathbb Z)$ invariant field strength in 10d is $\tilde F_5 = d C_4 + \frac{1}{2} B_2 \wedge F_3 - \frac{1}{2} C_2 \wedge H_3$ with $H_3 = d B_2$ and $F_3 = d C_2$, but since we are only interested in the gauging of $C_4$ these corrections play no role for us.} $F_5 = d C_4$. It is gauge invariant and satisfies the Bianchi identity $d F_5 = 0$.
Including the source terms, the relevant part of the action after taking the orientifold quotient becomes
\bea
S|_{C_4} &=& 2\pi\int  -\frac18  \, F_5\wedge \ast F_5+ \nonumber \\
 && 2 \pi \int C_4 \wedge   \left(   \frac12  \sum_{a} (Q_a({\bf F}_a) + Q_{a'}({\bf F}_{a'})  ) +   \frac{1}{2} \, Q({\bf R}) + \frac{1}{2} \sum_i (Q({\rm D3}_i) + Q({\rm D3}_{i'}) )  \right) \label{C4action} \,.
\eea
The source terms linear in $C_4$ follow by summing up the $C_4$ dependent contributions to the Chern-Simons action of the 7-branes, the O7-plane and the D3-branes listed in  Appendix \ref{app_conventions} as
\bea
Q_a({\bf F}_a)  &=& -\frac14 {\rm Tr} \, {\mathbf F}_a\wedge {\mathbf F}_a  \wedge [D_a] \label{QaFa} \\
Q({\bf R})   &=&  -  \frac{1}{16}   {\rm tr} \, {\mathbf R} \wedge {\mathbf R}  \wedge  [O7]   \\
Q({\rm D3}_i)    &=&   \frac{1}{2} [C_i] \,.
\eea 
Note the appearance of the trace ${\rm Tr}$, defined in (\ref{defTr}), in the expression (\ref{QaFa}). In the strict perturbative limit, in particular for gauge groups of type $SU(n)$, there is no difference compared to the trace in the fundamental representation. But more generally in F-theory, it is the object ${\rm Tr}$, rather than ${\rm tr}$, which appears in the Chern-Simons action.

As a result,  the Bianchi identity for  the field strength $F_5$ associated with $C_4$ now takes the non-standard form 
\be
\label{whole}
dF_5=  \frac{1}{2}  \sum_a  \left( {\rm Tr} \, \mathbf F_a\wedge \mathbf F_a  \wedge [D_a]+ {\rm Tr} \,  \mathbf F_{a'}\wedge \mathbf F_{a'} \wedge [D_{a'}] \right)  +   {\rm tr} \, {\mathbf R} \wedge {\mathbf R}  \wedge \frac{1}{8} [O7]     -  \sum_i ([C_i] + [C_{i'}]) \,.
\ee
To proceed further, we introduce the Chern-Simons forms $\mathbf{w}_{3a}$ for the gauge group on the 7-brane along $D_a$ as well as  $\mathbf{w}_{3Y}$ for the spin connection $\omega$ with the property
\bea
{\rm Tr} \, \mathbf F_a\wedge \mathbf F_a = d \, \mathbf{w}_{3a}, \qquad {\rm tr} \,  {\mathbf R} \wedge {\mathbf R} = d \, \mathbf{w}_{3Y} \,.
\eea
Similarly, one can define an Euler form $e_{5,i}$ associated with the 6-form $[C_i]$ Poincar\'e dual to the curve $C_i$ such that $d e_{5,i} = [C_i]$.\footnote{A careful definition can be found in \cite{Kim:2012wc}. A proper regularization of this term is necessary for a correct treatment of the normal bundle anomalies \cite{Cheung:1997az}, but this will play no role for us in this paper.}
This allows us to express  (\ref{whole}) as 
\bea \label{dF5a}
d\left( F_5 - \frac{1}{2} \sum_a \big( \mathbf{w}_{3a} \wedge [D_a] + \mathbf{w}_{3a'} \wedge [D_{a'}]\big)   -  \frac{1}{8}  \mathbf{w}_{3Y} \wedge [O7] + \sum_i  ( e_{5,i} +  e_{5,i'})    \right) = 0 \,,
\eea
which is solved by setting
\be \label{F5def}
F_5= d C_4  +  \frac{1}{2} \sum_a  \big(  \mathbf{w}_{3a}  \wedge [D_a]+\mathbf{w}_{3a'} \wedge [D_{a'}]   \big)
  + \frac{1}{8}  \mathbf{w}_{3Y} \wedge [O7] - \sum_i  ( e_{5,i} +  e_{5,i'}) \,  \,.
\ee
Taking into account the backreation of the source terms means that it is now this form of $F_5$ which appears in the kinetic term in (\ref{whole}). The full action (\ref{C4action}) is equivalent to
\bea
S|_{C_4} &=& 2\pi\int  -\frac18  \, F_5\wedge \ast F_5  +  \label{C4actionb} \\
 && 2 \pi \int F_5 \wedge   \left(   \frac18  \sum_{a} (\mathbf{w}_{3a} \wedge [D_a] + \mathbf{w}_{3a'} \wedge [D_{a'}] ) +   \frac{1}{32} \,\mathbf{w}_{3Y} \wedge [O7]   - \frac{1}{4} \sum_i (e_{5,i} + (e_{5,i'}) )  \right) \,,\nonumber
 \eea
 again with $F_5$ as in (\ref{F5def}).\footnote{Note that the cross-terms quadratic in the Chern-Simons terms vanish due their odd form degree.}

The form (\ref{F5def}) for the gauge invariant field strength $F_5$ implies that $C_4$ must transform non-trivially under gauge transformations associated with the 7-brane gauge group and the spin connection.
In absence of any background values for the fields, 
if under a gauge and Lorentz transformation the gauge connection $\mathbf{A}_a$ and the spin connection ${\bf \omega}$ change as
\bea
\mathbf{A}_a \rightarrow d \lambda_a + [\lambda_a,\mathbf{A}_a], \qquad \mathbf{\omega} \rightarrow d \chi + [\chi,{\bf \omega}] \,,
\eea
then the Chern-Simons forms vary as
\bea \label{deltawomega}
\delta \mathbf{w}_{3a} = d (\lambda_a  \,  d {\mathbf A}_a), \qquad \quad  \delta \mathbf{w}_{3Y} = d (\chi \, d {\bf \omega}) \,.
\eea
Since the field strength $F_5$ defined in (\ref{F5def}) is gauge invariant, this induces a corresponding gauge transformation of the potential $C_4$.
We are interested in situations in which both the gauge and the spin connection acquire non-trivial background values. 
Correspondingly we can 
decompose the field strength $\mathbf{F}$ into its fluctuation piece $F$ and a background component $\bar F$, and similarly for ${\bf R}$,
\bea \label{Fbarfdecomp}
\mathbf{F}=F+ \bar F, \qquad \quad \mathbf{R}=R+ \bar R \,.
\eea 
The gauge dependence of $C_4$ then becomes\footnote{Strictly speaking, we are not taking into account variations of the spin connection in the direction of the normal bundle, which are more subtle \cite{Cheung:1997az,Kim:2012wc} but play no role for us. Note also that, as we will argue momentarily, only abelian fluxes are of relevance for us so that we are writing $\bar F_a$ instead of $d \bar A_a$.}
 \bea\label{C4variation}
\delta_{\rm gauge} \, C_4 &=& -   \sum_{a} {\rm Tr} \, \lambda_{a}  \left(      (\bar F_a \wedge  [D_a]- \bar F_{a'} \wedge [D_{a'}]) +\frac{1}{2}(  d A_a \wedge [D_a]- d A_{a'} \wedge [D_{a'}]) \right)  \\
 \delta_{\rm spin} \, C_4 &=&    -  {\rm tr}  \, \chi {d \bar\omega} \wedge  \frac{1}{4} [O7]   -   {\rm tr} \, d \omega \wedge  \frac{1}{8}  [O7]   \label{Rterms} \,.
 \eea 
 Here we have used $\lambda_{a}=-\lambda_{a'}$, relating the gauge group on each brane along $D_a$ and its orientifold image. 
 The relative factor of $2$ in the first terms involving the background field strength and curvature results from expanding ${\mathbf F}_a^2 = 2 F_a \bar F_a + F_a^2 + \bar F_a^2$, and similarly for ${\bf R}$.
 As we will see next, the terms on the righthand side involving the internal background flux $\bar F_a$ induce the Green-Schwarz counterterms in the two-dimensional effective action, while the terms depending on the fluctuations  $F_a$ and $R$   contribute to the anomaly inflow counterterms for the anomaly from the 3-7 string modes.

\subsection{Derivation of the GS term in Type IIB}
In order to derive the Green-Schwarz counterterms, we first consider the  flux-dependent piece in the gauge variation of $C_4$, (\ref{C4variation}),
\be \label{deltaC4flux}
\delta_{\rm gauge} \, C_4 |_{{\rm flux}}=-  \sum_{a} {\rm Tr} \, \lambda_{a} \left( \bar F_a \wedge [D_{a}] - \bar F_{a'} \wedge [D_{a'}] \right) \,.
\ee
Due to the appearance of $C_4$ in the action (\ref{C4action}), while $F_5$ by itself is gauge invariant, this induces a gauge dependence of the effective action, which is precisely the manifestation of a Green-Schwarz counterterm. As we will see, the only relevant terms contributing to the Green-Schwarz terms are the couplings to $Q_a({\bf F}_a)$ and $Q_{a'}({\bf F}_{a'})$.
If we focus on these, substituting the variation (\ref{deltaC4flux})  of $C_4$ into \eqref{action} gives
\be
\ba
 \delta S_{\rm GS}=&\frac12 \left. \left(\sum_{b}\delta_{\rm gauge} \, S_b^{\rm D7}+\sum_{b'}\delta_{\rm gauge} \, S_{b'}^{\rm D7}\right) \right|_{\rm flux} \cr
=&\frac{2\pi}8 \int_{\mathbb{R}^{1,1}\times X_4}\sum_{a,b}  {\rm Tr}  \lambda_a   \left( \bar F_a\wedge [D_a] -    \bar F_{a'} \wedge [D_{a'}]   \right)    \wedge  \big( {\rm tr}  (\mathbf F_b\wedge \mathbf F_b)\wedge [D_b]+ {\rm tr}(\mathbf F_{b'} \wedge \mathbf F_{b'} )\wedge [D_{b'} ]  \big) \,, \cr
\ea
\ee
where we are indicating that after compactification the spacetime is of the form $\mathbb R^{1,1} \times X_4$. If we identify the fluctuations $F$ with the 2d field strength $F^{\rm 2d}$, we see that  
for  reasons of dimensionality only the last term in the decomposition 
\be
{\rm Tr}(\mathbf F_b\wedge \mathbf F_b)={\rm Tr}(F^{\rm 2d}_b\wedge F^{\rm 2d}_b)+{\rm Tr}(\bar F_b\wedge \bar F_b)+2 \, {\rm Tr}(F^{\rm 2d}_b\wedge \bar F_b)
\ee
makes a contribution. We thus find
\be
\delta S _{\rm GS}= \frac{2\pi}4\sum_{ab}  {\rm Tr}_a {\rm Tr}_b  \,  \lambda_a F_b^{\rm 2d}     \int_{X_4} \left((  \bar F_a \wedge[D_a]+\sigma^\ast(\bar F_a\wedge [D_a]) )\wedge ( \bar F_b \wedge[D_b]+\sigma^\ast( \bar F_b\wedge [D_b])) \right) \,.
\ee
Here we have used the definition  $\sigma^\ast({\rm Tr}\bar F_a\wedge [D_a])={\rm Tr} \bar F_{a'} \wedge [D_{a'}]$. Furthermore we have denoted the trace over the gauge group on brane $D_a$ with ${\rm Tr}_a$, and similarly for $D_b$.
Through the descent equations, this gauge variance yields the Green-Schwarz contribution to the anomaly polynomial
\be
\label{GSterm}
I_4^{\rm GS}=\frac14\sum_{a, b} {\rm Tr}_a {\rm Tr}_b F^{\rm 2d}_a\wedge F_b^{\rm 2d}\int_{X_4} \left( ( \bar F_a\wedge [D_a]+\sigma^\ast(\bar F_a \wedge[D_a])) \wedge (\bar F_b\wedge [D_b]+\sigma^\ast(\bar F_b\wedge [D_b])) \right) \,.
\ee

Note that the trace is taken simultaneously over the external and the internal components of the field strength, both for the gauge groups associated with $D_a$ and with $D_b$.
This implies that $I_4^{\rm GS}$ can only be non-vanishing for the abelian gauge symmetry factors in the two-dimensional effective action: Indeed, a contribution to a non-abelian gauge group would require at the same time non-abelian flux internally, but this would break the gauge group. The only option is that the flux is embedded along the direction of an abelian generator, which then acquires a Green-Schwarz anomaly term of the above form. 
This is a notable difference from the Green-Schwarz mechanism in six dimensions, which is well-known to operate also at the level of non-abelian gauge groups.

For a similar reason, the other source terms in (\ref{C4action}) do not contribute to the gauge variance of the classical action. Also, there can be no Green-Schwarz contribution to the pure  gravitational anomaly or even a mixed gauge-gravitational anomaly in two dimensions. 
This can be seen explicitly if one proceeds along the same lines with the background terms in (\ref{Rterms}) and uses the direct product structure of the Lorentz group as $SO(1,1) \times G_{\rm int}$ upon compactification.
In summary, the complete effect of the gauge dependence associated with the background term in (\ref{C4variation}) is the Green-Schwarz anomaly polynomial (\ref{GSterm}), while the background term in (\ref{Rterms}) does not lead to any gauge dependence of the effective action.


The Green-Schwarz counterterm (\ref{GSterm}) and in particular its overall normalization will be checked in a prototypical brane setup in Appendix \ref{app_AnomaliesIIB}, where we will verify that it correctly cancels the 1-loop anomalies induced by the 3-7 and the 7-7 sector.

\subsection{ 3-7 anomaly from gauging in Type IIB}

Let us now analyze the effect of the 
 dependent piece of the gauging (\ref{C4variation}),
\be \label{deltaCfluct}
\delta C_4|_{\rm fluct.}=- \frac{1}{2}  {\rm Tr} \sum_{a} \lambda_a(  d A^{\rm 2d}_a \wedge [D_{a}]-  d A^{\rm 2d}_{a'} \wedge [D_{a'}] )    -   {\rm tr} \, \chi \, { d \omega}^{\rm 2d} \wedge  \frac{1}{8}  [O7]  \,.
\ee
If we plug this expression into (\ref{C4action}) we receive  a contribution only from the internal components of the source terms.
Summing over all source terms associated with the D3-branes, the D7-branes and the O7-plane gives a vanishing total result because 
the total $C_4$ charge along the internal space $X_4$ vanishes as a result of D3-brane tadpole cancellation.
Nonetheless, each individual term by itself contains valuable information, namely (part of) the counterterms for the 1-loop gauge anomaly on the worldvolume of the respective branes.
By construction of the Chern-Simons brane actions, these counterterms \emph{locally} cancel the 1-loop anomaly associated with chiral modes on the worldvolume of the branes via the anomaly inflow mechanism \cite{Green:1996dd,Cheung:1997az,Minasian:1997mm}.
Tadpole cancellation then implies that the sum of all counterterms vanishes globally, which equivalent to the statement of anomaly cancellation. 


To extract the full anomaly inflow counterterm cancelling the 7-brane gauge anomalies from the 3-7 sector as well as the tangent bundle anomalies along the D3-brane, we follow the standard procedure \cite{Green:1996dd,Cheung:1997az,Minasian:1997mm} and rewrite the non-kinetic terms in the action (\ref{C4action})
as
\bea
S|_{C_4} &\supset &   S_1 + S_2 \\ 
S_1 & = & \frac{2 \pi}{4} \int  C_4 \wedge \sum_i ([C_i] + [C_{i'}] ) \\
S_2 &=& 2 \pi \int F_5 \wedge   \left(   \frac18  \sum_{a} (\mathbf{w}_{3a} \wedge [D_a] + \mathbf{w}_{3a'} \wedge [D_{a'}] ) +   \frac{1}{32} \,\mathbf{w}_{3Y} \wedge [O7] \right) \,.
 \eea

The anomaly inflow counterterms 
now have two contributions.
The first contribution comes from plugging the gauge variation (\ref{deltaCfluct}) into $S_1$,
\bea  \label{deltaS1inflow}
\left.\delta \, S_1 \right|_{\rm inflow} 
&=&- \frac{2\pi}8  \int_{\mathbb R^{1,1} } \sum_{a,i} {\rm Tr} \lambda_a   \, d A_a^{\rm 2d}  \int_{X_4}\big([D_a]+[D_{a'}])([C_i]+[C_{i'}]\big)  \\
& &    - \frac{2\pi}{32}   \int_{{\mathbb R}^{1,1}} {\rm tr} \, \chi \, d \omega^{\rm 2d}   \int_{X_4} [O7] \wedge \sum_{i} ([C_i] + [C_{i'}]) \label{deltaS1inflow-line2} \,,
\eea
where in the first line we have used that $A_{a'}^{\rm 2d}=-A_{a}^{\rm 2d}$.  
In addition, the Chern-Simons forms appearing in $S_2$ vary according to (\ref{deltawomega}).\footnote{We are here only taking into account the contribution to (\ref{deltawomega}) from the fluctuations of the fields; the contributions involving the background fields enter the Green-Schwarz terms and have hence already been taken into account in the previous section.}
After integration by parts we find a non-zero contribution because of the Bianchi identity (\ref{whole}). The relevant terms describing the anomaly inflow are obtained by plugging in only the last terms in (\ref{whole}), i.e. using $d F_5 =  - \sum_i ([C_i] + [C_{i'}]) + \ldots$. This gives a contribution of exactly the same form as (\ref{deltaS1inflow}) and hence altogether
\bea  \label{deltaS2inflow}
\left.\delta \, S \right|_{\rm inflow}    =  \left.\delta S_1 \right|_{\rm inflow}  + \left.\delta S_2 \right|_{\rm inflow}  = 2 \left.\delta S_1 \right|_{\rm inflow}  \,.
\eea
The terms (\ref{deltaS1inflow}) cancel the contribution to the 7-brane gauge group anomaly from the sector of 3-7 strings. 
By descent,  the associated 1-loop anomaly polynomial is therefore 
\be\label{anomaly37}
I_{4, \rm gauge}^{3-7}   =\frac{1}4   \sum_{a, i} {\rm Tr} F_a^{\rm 2d} \wedge F_a^{\rm 2d}  \int_{X_4} \left([D_a]+[D_{a'}])([C_i]+[C_{i'}] \right) \,.
\ee
Note that we have included a minus sign in $I_4^{3-7}$  because (\ref{deltaS2inflow}) represents the inflow counterterms to the actual 1-loop anomaly.
As the trace structure clearly shows, this contribution is non-vanishing also for simple gauge groups, in contrast to the Green-Schwarz terms derived earlier.

From the perspective of the effective  2d $(0,2)$ theory, the gauging (\ref{deltaCfluct}) translates into a gauging of the non-dynamical 2-forms obtained by dimensional reduction of $C_4$ in terms of internal 2-forms on $X_4$. This offers an interesting perspective on the contribution (\ref{anomaly37}) to the total anomaly polynomial: Rather than interpreting it as due to chiral localised defect modes we can view it as the effect of gauging these non-dynamical top-forms in the effective supergravity theory. 
This makes the formal similarity between the Green-Schwarz terms, associated with the gauging of the scalars from $C_4$, and the 3-7 anomaly on the righthand side of (\ref{gauge-flux1}) more natural.

The remaining terms (\ref{deltaS1inflow-line2}) cancel the contribution to the gravitational anomaly from all modes on the D3-brane worldvolume. This includes the 3-7 modes as well as the 3-brane bulk modes analyzed in detail in \cite{Lawrie:2016axq}. The associated anomaly polynomial is
\bea
I_{4, \rm grav}^{\rm D3} =  {\rm tr} \, R^{\rm 2d} \wedge R^{\rm 2d} \,  \int_{X_4}\frac{1}{16} [O7] \wedge \sum_{i} ([C_i] + [C_{i'}]) \,.
\eea

\subsection{F-theory lift} \label{sec_Ftheorylift}

It remains to uplift the perturbative results for the Green-Schwarz terms and the 3-7 anomaly to a description in fully-fledged F-theory, defined via duality to M-theory on an elliptic fibration $\hat X_5$. If a weakly coupled limit exists, the perturbative Type IIB Calabi-Yau $X_4$ is the double cover of the F-theory base $B_4$, with projection
\bea
\pi_+: X_4 \rightarrow B_4 \,. 
\eea
The cohomology classes even under the holomorphic involution $\sigma$ on $X_4$ uplift to cohomology classes of the same bidegree on $B_4$.
In particular, consider a divisor class $[D] \in H^{1,1}(X_4)$ and its image $\sigma^*[D]$ under the involution and define
\bea
[D_+] := [D] + \sigma_\ast[D] =: \pi^\ast_+ [D^{\rm b}]
\eea
with  $[D^{\rm b}] \in H^{1,1}(B_4)$. Then taking into account that $X_4$ is a double cover of $B_4$ the intersection numbers on both spaces are related as \cite{Krause:2012yh}
\bea \label{IIBtoF}
[D_{a+}] \cdot_{X_4} [D_{b+}] \cdot_{X_4} [D_{c+}] \cdot_{X_4} [D_{d+}] =  2 \, D^{\rm b}_a \cdot_{B_4} D^{\rm b}_b \cdot_{B_4} D^{\rm b}_c \cdot_{B_4} D^{\rm b}_d \,.
\eea

With this in mind consider first the perturbative expression (\ref{anomaly37}) for the 3-7 anomaly, with the aim of uplifting the sum over all brane stacks and their image to F-theory.
A divisor on $X_4$ wrapped by a non-abelian stack of 7-branes on $X_4$ uplifts, together with its image under the involution, to a corresponding divisor on $B_4$ according to the above rule, and this divisor on $B$ is a component of the discriminant locus carrying the corresponding non-abelian gauge group.
More subtle are the non-Cartan abelian gauge groups. In Type IIB language, $U(1)$ gauge symmetries which are massless in the absence of background flux are supported on linear combinations of divisors which are in the same class as their orientifold image.
Hence each abelian gauge group factor $U(1)_A$ is associated with a linear combination of (typically several) divisor classes $[D_a] + \sigma^\ast [D_a]$ on $X_4$.

Let us assume first that a brane configuration gives rise to no massless (in absence of fluxes) abelian gauge symmetries, i.e. the gauge group is only a product of non-abelian factors $G_I$, Then the uplift of  $\sum_{a} {\rm Tr} F_a^{\rm 2d} \wedge F_a^{\rm 2d}  ([D_a]+[D_{a'}])$ to F-theory is 
\bea
\sum_{i_I, j_J}   F^{\rm 2d}_{i_I} \wedge  F^{\rm 2d}_{j_J}  \,  {\rm Tr}\,  {\cal T}_{i_I} {\cal T}_{j_J}   D^{\rm b}_{I} = \sum_{i_I, j_J} F^{\rm 2d}_{i_I} \wedge  F^{\rm 2d}_{j_J} \,  \left(-\pi_\ast(E_{i_I} \cdot E_{j_J}) \right) \,.
\eea
Here we used (\ref{piEiEj}) to express the correctly normalised trace to $\pi_\ast(E_{i_I} \cdot E_{j_J})$.

In the presence of non-Cartan abelian symmetries, we must include these in the sum.
In F-theory language,  a non-Cartan gauge group factor $U(1)_A$ is generated by a 2-form $U_A$, defined via the Shioda-map as in (\ref{ShiodaUA}), but typically there is no separate component of the divisor $\Delta$ which we would associate with $U(1)_A$. This is because, form a 7-brane perspective, massless (in absence of gauge flux) $U(1)$s involve combinations of several divisor classes. However, the height-pairing (\ref{heightpairing}) defines a completely analogous object on the base $B_4$ including the information about the trace appearing in (\ref{anomaly37}).
Hence, the correct uplift of the expression for the 3-7 anomaly is 
\bea
\sum_{a} {\rm Tr} \,  F^{\rm 2d}_a \wedge F^{\rm 2d}_a \, ([D_a]+ [D_{a'}])   & \longrightarrow &  \sum_{\Lambda, \Sigma}  F^{\rm 2d}_\Lambda \wedge  F^{\rm 2d}_\Sigma     \, \left(-\pi_\ast(\mathfrak{F}_\Lambda \cdot \mathfrak{F}_\Sigma) \right)  \nonumber \\
\sum_{i} ([C_i] + [C_{i'}])    & \longrightarrow &    [C]  \\
\frac{1}{4} \sum_{a, i} {\rm Tr} F_a^{\rm 2d} \wedge F_a^{\rm 2d}  \int_{X_4} \left([D_a]+[D_{a'}])([C_i]+[C_{i'}] \right)  & \longrightarrow &   \frac{1}{2}   \, \sum_{\Lambda, \Sigma} F^{\rm 2d}_\Lambda \wedge  F^{\rm 2d}_\Sigma     \, \left(-\pi_\ast(\mathfrak{F}_\Lambda \cdot \mathfrak{F}_\Sigma) \right)   \cdot_{B_4}   [C] \nonumber
\eea
Here $[C]$ is the total class of the D3-brane on $B_4$ and we summing over all generators ${\mathfrak F}_\Sigma$, Cartan and non-Cartan. The last line in addition uses (\ref{IIBtoF}).
Hence 
\bea
I_{\rm 4, \rm gauge}^{3-7} = \sum_{\Lambda, \Sigma} F^{\rm 2d}_{\Lambda} F^{\rm 2d}_{\Sigma} \left(  \frac{1}{2}   \,      \, \left(-\pi_\ast(\mathfrak{F}_\Lambda \cdot \mathfrak{F}_\Sigma) \right) \, \cdot_{B_4}   [C] \right)
\eea 
 in precise agreement with our claim (\ref{37ABanomaly-b}) for the 3-7 gauge anomaly. 
 
 Note that in (\ref{anomaly37}), there appear no mixed anomaly contributions because in the perturbative limit the 3-7 strings can only be charged under the diagonal $U(1)_a$ gauge group of at most one D7-brane stack. On the other hand, if we sum over all massless (in absence of flux) $U(1)_A$ group factors (which are linear combinations of the $U(1)_a$ if a perturbative limit exists), mixed anomaly terms in general do result.

To uplift the 3-7 contribution to the gravitational anomaly polynomial, we recall from \cite{Krause:2012yh} the general rule that $\pi_+^\ast (c_{1}(B_4))= [O7]$, and therefore
\bea
\int_{X_4}  [O7] \wedge \sum_i ([C_i] + [C_{i'}])  & \longrightarrow&  2 c_1(B_4) \cdot_{B_4} [C] \,.
\eea 
The resulting expression 
\bea
I_{4, \rm grav}^{\rm D3} = \frac{1}{2} {\rm tr} \, R^{\rm 2d} \wedge R^{\rm 2d}    \left(  \frac{1}{4} c_1(B_4) \cdot_{B_4} [C] \right)\,.
 \eea
had already been derived in \cite{Lawrie:2016axq}.

It remains to uplift the Green-Schwarz anomaly polynomial (\ref{GSterm}) to F-theory.
Consider an internal  flux background associated with a line bundle whose structure group is identified with either a Cartan or a non-Cartan $U(1)$ subgroup. Such fluxes uplift in F-theory to expressions of the form $G_4 = \bar F \wedge {\mathfrak F}_{\Lambda}$ for the corresponding divisor generator that $U(1)$ symmetry. 
Employing once more (\ref{piEiEj}) and (\ref{heightpairing}), an expression of the form 
 ${\rm Tr}_a   \,  F^{\rm 2d}_a    ( \bar F_a \wedge[D_a]+\sigma^\ast(\bar F_a\wedge [D_a]) )$ uplifts to
 \bea
  F^{\rm 2d}_\Sigma ( - \pi_\ast(G_4 \cdot {\mathfrak F}_\Sigma)  ) \,.
 \eea
This remains correct even if the flux, on the Type IIB side, is associated with a $U(1)$ that is geometrically massive even before switching on the flux. Such fluxes lift to more general elements of $G_4$ \cite{Krause:2012yh,MayorgaPena:2017eda}.
Taking again into account the factor of $2$ from (\ref{IIBtoF}) we therefore arrive at
\bea \label{I4GS-F-theory}
I_4^{\rm GS} =   \sum_{\Lambda, \Sigma} F^{\rm 2d}_{\Lambda} F^{\rm 2d}_{\Sigma} \left( \frac{1}{2}      (  \pi_\ast(G_4 \cdot {\mathfrak F}_\Lambda)  ) \cdot_{B_4}    (  \pi_\ast(G_4 \cdot {\mathfrak F}_\Sigma)  )   \right)  
\eea
in agreement with our previous claim (\ref{GSclaim}). 

\subsection{Relation to 2d effective action}

For completeness, we can express our findings in the language of the 2d effective action and make contact with the formalism introduced in section \ref{sec_Anomaliesin2d}. 
Indeed, straightforward dimensional reduction of the action (\ref{C4action})  allows us to read off the kinetic metric $g_{\alpha \beta}$, the coupling matrix $\Omega_{\alpha \beta}$ and the gauging parameters $\Theta^\alpha_A$. This can be achieved by first performing the dimensional reduction in the language of Type IIB orientifolds and then uplifting the results according to the general rules described in section \ref{sec_Ftheorylift}.
We directly give the result in the language of F-theory: If we fix a basis $\omega_\alpha$ of $H^4(B_4,\mathbb R)$, the real scalar fields are obtained as 
\bea
C_4 = c^\alpha \, \omega_\alpha \,.
\eea
Matching the 10d and 2d kinetic terms in (\ref{C4action}) and (\ref{2daction1}), respectively, as well as the 10d self-duality condition $F_5 = \ast F_5$  with its 2d analogue (\ref{selfduality2d}) then fixes 
\bea
g_{\alpha \beta} &=& 2 \pi \int_{B_4} \omega_\alpha \wedge \ast \omega_\beta \\
 \Omega_{\alpha \beta} &=& 2 \pi \int_{B_4}  \omega_\alpha \wedge \omega_\beta =: 2 \pi \,  \tilde \Omega_{\alpha \beta} \,. 
\eea
Dimensional reduction of the interaction terms in (\ref{C4action}) finally identifies the gauging parameters
\bea
\label{gaugingterm1}
\Theta^\alpha_\Gamma = \tilde \Omega^{\alpha \beta} \int_{B_4}  \pi_\ast (G_4 \cdot \mathfrak{F}_\Gamma) \wedge \omega_{\beta}
\eea
in terms of the inverse matrix $\tilde \Omega$ satisfying  $\tilde \Omega^{\alpha \beta} \tilde \Omega_{\beta \gamma} = \delta^\alpha_{\, \,  \gamma}$.
As a check, plugging this expression into (\ref{factor}) correctly reproduces our result (\ref{I4GS-F-theory}) for the Green-Schwarz anomaly polynomial.

 \section{Example: $SU(5)\times U(1)$ gauge symmetry in F-theory} \label{sec_ExampleSU5U1}

In this section we exemplify our general expressions for the anomaly relations in an F-theory compactification on a Calabi-Yau 5-fold with gauge group 
$SU(5)\times U(1)$. The four-dimensional version of this model and its flux backgrounds has been studied in great detail in the literature \cite{Krause:2011xj,Krause:2012yh,Bies:2017fam,Bies:2017abs}, and its extension to Calabi-Yau five-folds has been discussed in \cite{Schafer-Nameki:2016cfr}. 
The geometry is sufficiently intricate to exemplify all interesting aspects of abelian, non-abelian and gravitational anomaly cancellation, while at the same time it avoids extra complications which arise when the codimension-two matter loci on the base $B_4$ are singular.

\subsection{Geometric background and 3-7 states} \label{sec_geomback}

We will briefly recall the properties of this model relevant for our discussion, referring for more details to \cite{Schafer-Nameki:2016cfr} as well as to \cite{Krause:2011xj,Bies:2017abs}, whose notation we adopt.

 We are considering the resolution of a Weierstrass model in Tate form defined by 
\be\label{Tate}
y^2 s e_3 e_4 + a_1  x y  z s+ a_{3,2}  y  z^3 e_0^2 e_1 e_4= x^3  s^2 e_1 e_2^2 e_3+ a_{2,1}  x^2 z^2 s e_0 e_1 e_2 +  a_{4,3} x z^4 e_0^3 e_1^2 e_2 e_4 \,,
\ee
where $[x: y : z]$ denote homogenous coordinates of the fibre ambient space $\mathbb P_{231}$ prior to resolution and $E_i: e_i=0$, $i=1,\ldots, 4$ represent the resolution divisors for the singularities associated with gauge group $SU(5)$. In addition to the zero section $S_0: z=0$, the fibration admits another independent rational section
 $S_A: s=0$. The resolved $SU(5)$ singularity sits in the fibre over the divisor $W: w=0$ on $B_4$, with $\pi^{-1}W: e_0 e_1 e_2 e_3 e_4 = 0$. With the help of Sage, we find the projection of $c_5(\hat X_5)$ and  $c_4(\hat X_5) $ of the resolved fibration ${\hat X_5}$ on the base $B_4$ and evaluate
\bea
\label{Chernclass}
\pi_\ast (c_5(\hat X_5))&=&-576c_1^4+1464c_1^3 W-48 c_1^2 c_2-1410 c_1^2 W^2+46 c_1 c_2 W+12 c_1c_3+608c_1 W^3 \nonumber \\
&& -18 c_2 W^2-102 W^4 \label{c5hatX5} \\
\pi_\ast (c_4(\hat X_5))&=&144 c_1^3-264 c_1^2 W+12  c_1 c_2+162 {c_1} W^2-30 W^3 \,. \label{c4hatX5}
\eea
 Here and in the sequel, the Chern classes $c_i$ without any specification denote $c_i(B_4)$ and all of the intersection numbers between the divisors are evaluated on $B_4$.
 Finally, $a_{i,j}$ define the following divisor classes on the base $B_4$ with $c_1(B_4) =: c_1$,
\be
\label{classtatecoefficient}
[a_1]=c_1\,,\quad 
[a_{2,1}]= 2 c_1-W \,,\quad 
[a_{3,2}]=3 c_1-2 W\,,\quad 
[a_{4,3}]= 4 c_1-3 W\,.\quad 
\ee
The discriminant of the blowdown of this model (setting $e_i = 1$ for $i=1, \ldots, 4$) is
\be
\Delta= w^5 \left( a_1^4 a_{3,2} \left(a_{2,1} a_{3,2}-a_1 a_{4,3}\right) w^5  + O(w) \right) 
\ee
and indicates that there are four codimension-two matter loci on $B_4$ with classes 
\be
\label{matterloci}
\ba
C_{{\bf 10}_{1}}:\qquad  &W \cdot [a_1]=  W \cdot c_1\cr 
C_{{\bf 5}_{3}}:\qquad& W \cdot [a_{3,2}]= W \cdot (3 c_1-2 W)\cr
C_{{\bf 5}_{-2}}:\qquad &  W \cdot[a_1 a_{4,3} - a_{2,1} a_{3,2}] = W \cdot (5 c_1-3 W) \cr
C_{{\bf 1}_{5}}: \qquad & [a_3]\cdot[a_{4,3}] = (3 c_1 -2 W) \cdot (4 c_1 -3 W) \,.
\ea
\ee
The subscripts denote the charges under the non-Cartan $U(1)_A$ associated with the divisor \cite{Krause:2011xj}\footnote{We are using the conventions of \cite{Bies:2017fam,Bies:2017abs}, where in particular the fibre structure and the resulting charge assignments are detailed.}
\be \label{noncartan}
U_A =- \left( 5 (S_A - S_0 - c_1) + 2 E_1 + 4 E_2 + 6 E_3 + 3 E_4 \right) \,.
\ee
Note that in this example all of the codimension-two loci are smooth, while in principle they could exhibit singularities. In this case the chirality formula \eqref{chiS2} would receive corrections \cite{Schafer-Nameki:2016cfr}.
The height pairing associated with $U_A$ is 
\bea \label{DA}
D_A = - \pi_* (U_A \cdot U_A) = - 30 W + 50 c_1 \,.
\eea

The D3-brane tadpole requires the inclusion of D3-branes wrapping a curve of total class $C$ constrained as in (\ref{Cclass}).
In the present example, each intersection point between $C$ and the $SU(5)$ divisor $W$ carries one Fermi multiplet in the fundamental representation ${\bf 5}_{q_1}$ of $SU(5)$ \cite{Schafer-Nameki:2016cfr} with $U(1)_A$ charge $q_1$. 
The intersections with the remainder of the discriminant  carry additional Fermi multiplets, whose determination is very subtle due to $SL(2, \mathbb Z)$ monodromies along $C$. In general some of these will have a non-zero $U(1)_A$ charge, while some may be completely uncharged under $SU(5) \times U(1)_A$.
Our knowledge of the net contribution (\ref{37ABanomaly}) of the 3-7 sector to the abelian anomaly together with its contribution to the gravitational anomaly \cite{Lawrie:2016rqe} allow us to constrain this matter as follows:
Let us adopt from the discussion around (\ref{D37notation}) the notation $D_{37}(\bf R)$ for the divisor class on $B_4$ such that the effective chiral index of 3-7 states in representation ${\bf R}$ is given by $\chi({\bf R}) = - [C] \cdot D_{37}({\bf R})$. Then  $D_{37}({\bf 5}_{q_1}) = W$, and the remaining divisor classes are constrained by the abelian and gravitational anomaly as 

\bea
\label{3-7locis1}
 5 \, q_1^2 \,  D_{37}({\bf 5}_{q_1}) + \sum_i  q_i^2 \, D_{37}({\bf 1}_{q_i}) &=&  D_A =  {-30 W + 50 c_1 } \\
 5 \, D_{37}({\bf 5}_{q_1}) + \sum_i  D_{37}({\bf 1}_{q_i}) &=& 8 c_1 \,.
\eea
These equations are consistent with the assertion that, in addition to the states ${\bf 5}_{q_1}$, there is only one further type of 3-7 Fermi multiplets in representation ${\bf 1}_{q_2}$ with charge assignments
\bea \label{chargesU1A}
|q_1| = \frac{1}{2}, \qquad \quad |q_2| = \frac{5}{2} 
\eea
such that 
\bea \label{D37152}
 D_{37}({\bf 1}_{q_2}) = - 5 W + 8 c_1.
 \eea 
These values are in complete agreement with the perturbative limit of the compactification: 
To see this, recall from \cite{Krause:2012yh} that the Type IIB limit consists of a brane stack (plus image) with gauge group $U(5)_a$ and another brane-image brane pair  carrying gauge group $U(1)_b$.  The geometrically massless $U(1)$ symmetry is given by the linear combination $U(1)_A = \frac{1}{2} (U(1)_a - 5 U(1)_b)$, where $U(1)_a$ is the diagonal $U(1)$ of $U(5)_a$ (cf. equ. 4.3 of \cite{Krause:2012yh}) and the normalization conforms with the definition (\ref{noncartan}) of the $U(1)_A$ generator. The 3-7 modes at the intersection of $C$ with the $U(5)_a$ stack hence carry charge $|q_1| = \frac{1}{2} |(1 + 0)|$ and transform as ${\bf 5}$ of $SU(5)_a$ and those at the intersection of $C$ with $U(1)_b$ are $SU(5)_a$ singlets with charge $|q_2| = \frac{1}{2}| (0 - 5) |$. The class (\ref{D37152}) furthermore coincides with the class of the $U(1)_b$ brane as dictated by the 7-brane tadpole cancellation condition.

We stress  that more generally the pattern of singlets in the 3-7 sector can be more intricate. What is uniquely determined, however, is the net contribution of the 3-7 states both to the gauge and the gravitational anomalies.

Now we are in the position to check our proposal \eqref{AnomaliesF-theory} within this example. As we have discussed before, we expect that the curvature and the flux induced anomalies should each cancel among themselves. Therefore, in the following we split our proof into three parts: We begin with the flux independent contribution to the anomalies and verify their precise cancellation as a result of rather sophisticated relations between the topological invariants of the resolved 5-fold.
Next we consider the two different types of $G_4$ flux spanning the space of fluxes within $H^{2,2}_{\rm vert}(\hat X_5)$ with the purpose of verifying in particular our proposal for the Green-Schwarz term \eqref{gaugeanomalyFtheory}, and it will be shown that the anomalies induced by the two $G_4$ fluxes are cancelled very neatly.

\subsection{Curvature dependent anomaly relations}
 
In this section, we verify that the conditions  \eqref{AnomaliesF-theory} for anomaly cancellation are satisfied in the absence of background flux, i.e. $G_4 = 0$.  This amounts to evaluating (\ref{gauge-geom1}) for the gauge and (\ref{curvgravanomalies}) for the gravitational anomalies.

The various 7-brane codimension-two matter loci $C_{\bf R}$ have been listed in \eqref{matterloci}, and in the present example they are all smooth \cite{Schafer-Nameki:2016cfr} such that the index theorem can be applied as in (\ref{chigeom1}).
With the help of the adjunction formula we find the following flux independent part of the chiral indices for the matter surfaces,
\be
\ba
\chi({\bf 10}_{1})|_{\rm geom} &= \frac{1}{24} c_1 W \left(2 c_2+W^2\right)\cr 
\chi({\bf 5}_{3})|_{\rm geom} &= \frac{1}{24} W \left(3 c_1-2 W\right) \left(-12 c_1 W+8 c_1^2+2 c_2+5
   W^2\right)\cr 
\chi({\bf 5}_{-2})|_{\rm geom} &=   \frac{1}{12} W \left(5 c_1-3 W\right) \left(-15 c_1 W+12 c_1^2+c_2+5 W^2\right) \cr
\chi({\bf 1}_{5})|_{\rm geom} & =    \frac{1}{24} \left(4 c_1 - 3 W\right) \left(3 c_1 - 2 W\right) \left(24 c_1^2 + 2 c_2 - 36 c_1 W + 13 W^2\right) .
\ea
\ee

 The first equation in \eqref{AnomaliesF-theory}, i.e. the purely non-abelian $SU(5)$ gauge anomaly, has been verified in \cite{Schafer-Nameki:2016cfr}.  For this analysis to be self-contained, let us briefly recap the computation as a warmup. With the appropriate anomaly coefficients (\ref{cr2}), $c^{(2)}_{\bf 10} = {3}$, $c^{(2)}_{\bf 5}= 1$, the matter from the 7-brane codimension-two loci contributes to the non-abelian anomaly (\ref{AI-gen1})
\be
{\cal A}_{SU(5)}|_{\rm surface, geom}= {\frac 3 2}  \chi({{\bf 10}_{1}})|_{\rm geom} + {\frac1 2}\chi({{\bf 5}_{3}})|_{\rm geom} + {\frac 1 2}\chi({\bf 5}_{-2})|_{\rm geom} \,.
\ee
The chiral matter from the 7-brane bulk transforms  in the adjoint with  $c^{(2)}_{\bf 24} =10$ and contributes
\be
\mathcal{A}_{SU(5)}|_{\rm bulk, geom}=  5   \chi({\bf 24}_0)|_{\rm geom} = -\frac{5}{24} W \left(c_1-W\right) \left(W \left(W-c_1\right)+c_2\right) \,,
\ee
where we have used (\ref{chibadj1}).
In addition, there is another contribution from anti-chiral fermions generated in the 3-7 sector. These modes transform  in representation ${\bf 5}_{q_1}$ and their chiral index is given by minus the point-wise intersection number $ -  [W]\cdot [C]$ with $[C]=\frac1{24} \pi_\ast(c_4(\hat X_5)$ in the absence of flux. 
With the help of (\ref{c4hatX5}), their $SU(5)$ anomaly contribution follows as
\bea
\mathcal{A}_{SU(5)}|_{3-7, {\rm geom}}    &=&\frac12\chi_{3-7}(\mathbf{5}_{q_1}) |_{\rm geom} =-\frac12 W\cdot \frac1{24} \pi_\ast(c_4(\hat X_5))\cr
&=&-\frac1{48}W\cdot (144c_1^3-264c_1^2W+12c_1c_2+162c_1W^2-30W) \,.
\eea
Then the pure non-abelian $SU(5)$ anomalies, in the absence of $G_4$ fluxes, indeed cancel,
\be
\mathcal{A}_{SU(5)}|_{3-7, {\rm geom}} + \mathcal{A}_{SU(5)}|_{{\rm bulk, geom}}   + {\cal A}_{SU(5)}|_{\rm surface, geom} =0 \,.
\ee

Now we switch gear to check the cancellation of the $U(1)_A$ gauge anomaly.  As we have discussed above, there are two types of charged matter states in the 3-7 sector with different $U(1)_A$ charges. With the help of \eqref{3-7locis1}, their combined contribution to the abelian anomalies is
\be
\mathcal{A}_{U(1)}|_{{3-7},\rm geom} = \frac{5}{2} q_1^2 \, \chi_{3-7}(\mathbf{5}_{q_1})|_{\rm geom}  + \frac{1}{2}   q_2^2 \chi_{3-7}({\bf 1}_{q_2})|_{\rm geom} =- \,  \frac1{48} \pi_\ast (c_4(\hat X_5)) \cdot {(-30W+50 c_1)} \,.
\ee
This  perfectly cancels the anomalies from the 7-7 sector,
\bea
\mathcal{A}_{U(1)}|_{\rm geom} &=&\frac{1}{2}\sum_{\bf R} q^2_{A}({\bf R})   {\rm dim}({\bf R})   \chi({\bf R})|_{\rm geom} \cr
&=&\frac{1}{2} \left[ 10\chi(\mathbf{10}_{1})+20\chi(\mathbf{5}_{-2})
+45 \chi(\mathbf{5}_{3})+25\chi(\mathbf{1}_5)  
+( 5 q_1^2 \, \chi_{3-7}(\mathbf{5}_{q_1})  +   q_2^2 \chi_{3-7}({\bf 1}_{q_2})) \right]|_{\rm geom}\cr
&=& 0
\eea
as it must since the Green-Schwarz counterterms vanish in absence of flux.

 Finally, let us compute the gravitational anomalies. In absence of flux, gravitational anomaly cancellation is equivalent to (\ref{curvgravanomalies}) over a generic base $B_4$.
This equation involves the Chern class $c_5(\hat X_5)$ and $c_4(\hat X_5)$ of the resolved Calabi-Yau five-fold $\hat X_5$.
 With the help of \eqref{Chernclass}, we find
\bea
\Delta\chi_1(\hat X_5)&=&-66 c_1^4-61  c_1^3 W- {c_1}^2  {c_2}+\frac{235  c_1^2 W^2}{4}-\frac{23 c_1 c_2 W}{12}-\frac{76c_1 W^3}{3}+\frac{3 c_2 W^2}{4}+\frac{17 W^4}{4} \cr
-6c_1\Delta [C] &=&54 c_1^4+66 c_1^3 W-\frac{81 c_1^2 W^2}{2}+\frac{15  c_1 W^3}{2}  \,.
\eea
Summing both terms up perfectly matches the RHS of (\ref{curvgravanomalies}),
\bea
&& - (10 \chi(\mathbf{10}_{1})+5\chi(\mathbf{5}_{-2})+5\chi(\mathbf{5}_{3})+\chi(\mathbf{1}_5) +24  \chi({\bf 24}_0)-4\chi({\bf 24}_0))|_{\rm geom} \cr
&=& - (12c_1^4-5 c_1^3 W+c_1^2 c_2-\frac{73c_1^2 W^2}{4}+\frac{23c_1c_2 W}{12}+\frac{107c_1W^3}{6}-\frac{3 c_2 W^2}{4}-\frac{17 W^4}{4}) \,.
\eea

In summary, we have checked that in this example with the absence of $G_4$ fluxes, all types of anomalies are cancelled by themselves and in agreement with \eqref{AnomaliesF-theory}.

\subsection{Flux dependent anomaly relations} \label{subsec_fluxdepan}

In the  $SU(5) \times U(1)_A$ model defined by (\ref{Tate}), there only exist two types of gauge invariant 4-form fluxes $G_4 \in H^{2,2}_{\rm vert}(\hat X_5)$ compatible with the $SU(5) \times U(1)_A$ gauge group \cite{Krause:2012yh}.  We choose a basis of fluxes as 
\bea \label{fluxes-normalisation}
G_4^A &=& F \cdot [U_A] \label{G4A}\\
G_4^\lambda &=& - \lambda \left(   E_2 \cdot E_4 + \frac{1}{5} (2 E_1 - E_2  + E_3 -2 E_4) \cdot c_1  \right) \,.
\eea
Here $[U_A]$ is the 2-form class dual to the non-Cartan $U(1)_A$ divisor $U_A$ defined in (\ref{noncartan}), $\lambda$ is a constant and $F \in H^{1,1}(B_4)$ is an arbitrary class parametrizing the flux. Both $\lambda$ and $F$ are to be chosen such that $G_4 + \frac{1}{2} c_2(\hat X_5) \in H^2(\hat X_5,\mathbb Z)$.
We now analyze the anomaly relations, including the Green-Schwarz terms, for both of these flux backgrounds in turn.

\subsubsection{$G_4^A$ flux}

We begin with the flux background (\ref{G4A}). 
The cancellation of non-abelian $SU(5)$ gauge anomalies in the presence of $G_4^A$ has already been verified in \cite{Schafer-Nameki:2016cfr} so that we can focus on
evaluating \eqref{gauge-flux1}, or equivalently \eqref{Abelian1}, for the $U(1)_A$ anomaly.
To compute the flux dependent chiral index of the 7-brane various matter states, we need to extract the line bundle $L_{\bf R}$ defined in (\ref{c1LR}) on the 7-brane codimension-two matter loci. Since $G_4^A$  is simply the gauge flux associated with the non-Cartan factor $U(1)_A$, we know that $\pi_\ast(G_4^A \cdot S^a_{\bf R}) = q_A({\bf R}) F|_{C_{\bf R}}$. It follows that 
\bea
c_1(L_{{\bf 10}_1}) = \left. F \right|_{C_{\mathbf{10}_{1}}}, \qquad c_1(L_{{\bf 5}_3}) = 3\left. F \right|_{C_{\mathbf{5}_{3}}}, \qquad c_1(L_{{\bf 5}_{-2}}) = \left. -2 F \right|_{C_{\mathbf{5}_{-2}}}, \qquad c_1(L_{{\bf 1}_5}) =  5\left. F \right|_{C_{\mathbf{1}_{5}}} 
\eea
and therefore 
\bea
\chi( \mathbf{10}_{1})|_{\rm flux}&=& \frac12 \int_{C_{{\mathbf{10}}_1}} F^2, \qquad \chi(\mathbf{5}_{-2})|_{\rm flux}=\frac12 \int_{C_{\mathbf {5}_{-2}}}(-2F)^2 \\
 \chi({\mathbf 5}_{3})|_{\rm flux}  &=&  \frac12 \int_{C_{{\bf 5}_{3}}}(3F)^2,\qquad  \chi(\mathbf{1}_{5})|_{\rm flux}  =  \frac12 \int_{C_{{\bf 1}_{5}}}(5F)^2  .
\eea
The anomaly contribution (\ref{37ABanomaly-b}) from the 3-7-brane sector is 
\bea
{\cal A}_{U(1)}|_{3-7,{\rm flux}} =  - \frac{1}{4} F \cdot F \cdot_{B_4}  \pi_\ast([U_A] \cdot [U_A])  \cdot_{B_4} \pi_\ast([U_A] \cdot [U_A])
\eea
with  $\pi_\ast([U_A] \cdot [U_A])  = - D_A$ as in (\ref{DA}).
Altogether this gives for the LHS of \eqref{Abelian1} 
\bea
\mathcal{A}_{U(1)}|_{\rm flux}&=&  \mathcal{A}_{U(1)}|_{7-7,{\rm flux}} +   {\cal A}_{U(1)}|_{3-7,{\rm flux}}    \cr
&=&\frac12 \left(10\chi(\mathbf{10}_{1})|_{\rm flux}+20\chi(\mathbf{5}_{-2})|_{\rm flux}
+45 \chi(\mathbf{5}_{3})|_{\rm flux} +25 \chi(\mathbf{1}_{5})|_{\rm flux}  \right)        - \frac{1}{4} F^2 D_A^2  \\
&=& \frac12 F^2 (50 c_1 - 30 W)^2 \,.
\eea
This is to be compared to the RHS of \eqref{Abelian1} given by the Green-Schwarz counterterms (\ref{GSclaim})
 \bea
\frac1{4\pi}  \Omega_{\alpha \beta}   \Theta_A^\alpha  \Theta_B^\beta  &=&\frac12 \pi_{\ast} (G_4 \cdot G_4) \cdot_{B_4}  \pi_\ast ([U_A] \cdot [U_A]) =\frac12 F \cdot_{B_4} F \cdot_{B_4} \pi_\ast([U_A] \cdot [U_A])^2 \cr
&=& \frac12 F^2 \cdot (50 c_1 - 30 W)^2 \,.
\eea
Hence \eqref{Abelian1}  and therefore \eqref{gauge-flux1} hold.

 Finally,  let us switch to cancellation of the purely gravitational anomaly. Given the above expressions, the LHS of \eqref{gravitationalanomalies2} yields 
\be
-6 c_1 \cdot \pi_{\ast}(G_4^A \cdot G_4^A)  = - 6 c_1 \cdot F \cdot F \cdot (-D_A) =         -6 c_1 F^2 (-50 c_1 +30 W)
\ee
which perfectly matches  the RHS of \eqref{gravitationalanomalies2} given by
\be
2 \left(10\chi(\mathbf{10}_{1})|_{\rm flux}+5\chi({\mathbf 5}_{3})|_{\rm flux}+5\chi({\mathbf 5}_{-2})|_{\rm flux}+ \chi(\mathbf{1}_{5})|_{\rm flux} \right)=6 c_1 F^2 (50 c_1-30 W) \,.
\ee

\subsubsection{$G_4^\lambda$ flux}

Verifying the anomalies in the presence of flux of the form $G_4^\lambda$ is slightly more involved. In the sequel we heavily build on the analysis of \cite{Bies:2017fam}, where this gauge background is described, in a compactification to four dimensions, as a 'matter surface flux'. Since the fiber structure is the same, we can extend these results to F-theory compactification on an elliptic 5-fold. Since we are now working over a base of complex dimension four, extra technical complications arise in the computation of the chiral index for the 7-brane ammeter, which we will solve in appendix \ref{app_chirality}.

Key to computing the 7-brane matter chiralities is again the induced line bundle $L_{\bf R} = \pi_\ast(G_4^\lambda \cdot S^a_{\bf R})$, given this time by 
\bea
c_1(L_{{\bf 10}_1})& =& \frac{-3\lambda}{5} [Y_1] + \frac{4 \lambda}{5} [Y_2], \qquad c_1(L_{{\bf 5}_3}) = \frac{-2\lambda}{5} [Y_2], \\
 c_1(L_{{\bf 5}_{-2}}) &=& \frac{3 \lambda}{5} [Y_1] - \frac{2 \lambda}{5} [Y_2], \qquad 
c_1(L_{{\bf 1}_5}) =  0 \,.
\eea
A derivation can be found in section 5 of \cite{Bies:2017fam}. By Poincar\'e duality,  the objects $Y_i$ describe curve classes on the respective matter codimension-two loci on the base, defined as the intersection loci
\begin{align}
\begin{split} \label{transverseY}
  C_{\mathbf{5}_{3}} \cap C_{\mathbf{10}_{1}}&= Y_2 \, , \\
  C_{\mathbf{5}_{-2}} \cap C_{\mathbf{10}_{1}} &= Y_1 + Y_2 \, , \\
  C_{\mathbf{5}_{-2}} \cap C_{\mathbf{5}_{3}} &= Y_2 + Y_3  \, .
\end{split}
\end{align}
The first Chern classes of the line bundles $L_{{\bf 10}_1}$ and $L_{{\bf 5}_3}$  can be expressed as the pullback of divisor classes from $W$ to the respective matter loci,
\bea
c_1(L_{\mathbf{10}_{1}}) &=&  \frac{\lambda}{5}   \left(  -3 ([Y_2] + [Y_1])   + 7 [Y_2]   \right)|_{C_{\mathbf{10}_{1}}} =  \frac{\lambda}{5}   \left(  6 c_1 - 5 W  \right)|_{C_{\mathbf{10}_{1}}} \label{L101lambdabundle}\\
c_1(L_{\mathbf{5}_{3}})   &=&\frac{\lambda}{5}   \left(    - 2 [Y_2]   \right)|_{C_{\mathbf{5}_{3}}}     =  \frac{\lambda}{5}   \left(    - 2 c_1   \right)|_{C_{\mathbf{5}_{3}}} \,.
\eea
Hence we can straightforwardly compute the associated chiralities  as integrals on $B_4$
\bea
\chi(C_{\mathbf{10}_{1}})|_{\rm flux}  &=& \frac{1}{2} \int_{C_{\mathbf{10}_{1}}}   c_1^2(L_{\mathbf{10}_{1}})  = \frac{\lambda^2}{50}\,   W \cdot c_1 \cdot (  6 c_1 - 5 W  )^2  \,, \\
\chi(C_{\mathbf{5}_{3}})|_{\rm flux} &=&   \frac{1}{2} \int_{C_{\mathbf{5}_{3}}}   c_1^2(L_{\mathbf{5}_{3}})    = \frac{\lambda^2}{50}   \,  W \cdot (3 c_1 - 2 W)  \cdot 4 c_1^2 \,.
\eea
By contrast, $c_1(L_{{\bf 5}_{-2}})$ cannot be interpreted as the class of a complete intersection of a base divisor with  $C_{\mathbf{5}_{-2}}$ \cite{Bies:2017fam}.
Each of the classes $Y_i$ defines a divisor class on $C_{{\bf 5}_{-2}}$, dual to a curve. 
The technical difficulty is that $Y_1$ and $Y_2$ separately cannot be written as the pullback of a divisor class from the 7-brane divisor $W$ to $C_{{\bf 5}_{-2}}$.
Rather, on $W$, the curves $Y_i$ are given by intersections
\bea \label{defYi}
Y_1 = a_1  \cap a_{2,1}|_W, \qquad  Y_2 = a_1 \cap a_{3,2}|_W, \qquad Y_3 = a_{4,3} \cap a_{3,2}|_W \,,
\eea
where the class of these Tate coefficients have been listed in \eqref{classtatecoefficient}.
In appendix \ref{app_chirality} we will discuss how to evaluate the chirality of ${\bf 5}_{-2}$ despite this complication, our  final result being
\be \label{chi5-2lambda}
\chi(C_{\mathbf{5}_{-2}})|_{\rm flux}  =   \frac{1}{2} \int_{C_{\mathbf{5}_{-2}}}   c_1^2(L_{\mathbf{5}_{-2}})   =  -\frac{\lambda^2}{25}c_1 \cdot  W \cdot \left(60 c_1^2-79 c_1W+25 W^2\right) \,.
\ee

In light of the discussion of section (\ref{sec_geomback}), the chiral indices for the 3-7 matter states as induced by $G_4^\lambda$ take the form
\bea \label{37lambda1}
\chi_{3-7}(\mathbf{5}_{q_1}) & =&  - [C]|_{\rm flux} \cdot W  \\
\chi_{3-7}({\bf 1}_{q_2}) & =&   -[C]|_{\rm flux} \cdot (- 5 W + 8 c_1)     \,,
\eea
where the flux dependent piece of the 3-brane class reads
\bea \label{G4double}
[C]|_{\rm flux} = - \frac{1}{2} \pi_\ast(G_4^\lambda \cdot G_4^\lambda) = - \frac{\lambda^2}{10} W \cdot c_1 \cdot (6 c_1 - 5 W) \,.
\eea
To derive this latter result, recall from section 4.3 of \cite{Bies:2017fam} that up to irrelevant correction terms $G_4^\lambda$ for $\lambda=1$ is the class associated with one of the matter fibrations $S^a_{{\bf 10}_1}$. The result for $\pi_\ast(G_4^\lambda \cdot G_4^\lambda)  =  \lambda \,  \pi_\ast( G_4^\lambda \cdot S^a_{{\bf 10}_1}$ can then be read off from (\ref{L101lambdabundle}).

We are finally in a  position to check the cancellation of anomalies in the presence of $G_4^\lambda$, beginning with the pure non-abelian gauge anomaly. Note the $G_4^\lambda$ background does not induce any chirality for the 7-brane bulk matter. Together with the above explicit expressions for chiral indices in the 7-brane and the 3-7 sector, one can easily confirm that 
 \bea
&&\mathcal{A}_{SU(5)}|_{\rm flux}= \frac32 \chi(C_{\mathbf{10}_{1}})|_{\rm flux} +\frac12\chi(C_{\mathbf{5}_{3}})|_{\rm flux} +\frac12 \chi(C_{\mathbf{5}_{-2}})|_{\rm flux} +\frac12 \chi_{3-7}(\mathbf{5}_{q_1})|_{\rm flux}=0 \,.
 \eea

Next we turn to the $G_4^\lambda$ dependent part of the abelian gauge anomalies.
The combined 1-loop anomaly from the 7-7 and the 3-7 matter evaluates to
\bea \label{U1lambdaanomaly}
\mathcal{A}_{U(1)_A}|_{\rm flux}&=&\frac{1}{2}\sum_{\bf R} \text{dim}({\bf R})   \,  q_A^2({\bf R})  \,  \chi({\bf R})|_{\rm flux}   \cr
&=&\frac12 \left( 10\chi(\mathbf{10}_{1}) +20\chi(\mathbf{5}_{-2})  
+45 \chi(\mathbf{5}_{3}) +25\chi(\mathbf{1}_{5}) + 5 q_1^2 \, \chi_{3-7}(\mathbf{5}_{q_1})    +   q_2^2 \chi_{3-7}({\bf 1}_{q_2})  \right) |_{\rm flux} \cr
&=&\frac{1}{2} \lambda^2 \, c_1^2  \, W^2 \,.
\eea
For the 3-7 contribution we can either use (\ref{37lambda1}) with the charge assignments (\ref{chargesU1A}), or directly evaluate the $G_4^\lambda$ dependent component of   (\ref{37ABanomaly}). 
The combined 1-loop anomaly forms the LHS of (\ref{Abelian1}) and must be cancelled by the Green-Schwarz terms \eqref{GSclaim} appearing on the RHS.
To compute the latter, we make again use of the interpretation of $G_4^\lambda$ as one of the matter fibrations $S^a({\bf 10}_1)$. Intersection this with the $U(1)_A$ generator $U_A$ in the fiber reproduces the $U(1)_A$ charge of ${\bf 10}_1$ and therefore
\bea
\pi_\ast (G_4 \cdot [U_A] )= \lambda \,  C_{{\bf 10}_1} \cdot W   =   \lambda \, c_1 \cdot W \,.
\eea
With this the Green-Schwarz terms are 
\bea
\label{GSterm2}
&& \frac1{4\pi}  \Omega_{\alpha \beta}   \Theta_A^\alpha  \Theta_B^\beta =\frac12  \pi_\ast(G_4^\lambda \cdot [U_A])  \cdot \pi_\ast(G_4^\lambda \cdot [U_A])  =\frac12\lambda^2 \, c^2_1 \cdot  W^2 \,.
\eea
This perfectly cancels the 1-loop anomalies (\ref{U1lambdaanomaly}) and hence verifies the $G_4^\lambda$ dependent part of (\ref{Abelian1}) or equivalently (\ref{gauge-flux1}).

As for the cancellation of the  gravitational anomalies, with the help of \eqref{G4double},  the LHS of \eqref{gravitationalanomalies2} becomes
\be
 -6 c_1 \cdot \pi_\ast (G_4^\lambda \cdot  G_4^\lambda)  = - \frac{6}{5}  \lambda^2 \,  c_1^2 \cdot  W \cdot (6 c_1-5 W),
\ee
which is again exactly equal to the RHS of \eqref{gravitationalanomalies2} 
\be
2 \,  (10\chi(\mathbf{10}_{1})+5\chi({\mathbf 5}_{3}) +5\chi({\mathbf 5}_{-2}) + \chi(\mathbf{1}_{5}) )|_{\rm flux}=-\frac{6}{5}  \lambda^2 \,  c_1^2 \cdot  W \cdot (6 c_1-5 W) \,.\ee

\section{Comparison to 6d and 4d anomaly relations} \label{sec_Comparison6d4d}

In this final section we 
compare the 2d anomaly relations (\ref{gaugeanomalyFtheory}) and (\ref{gravitationalanomalies1}) to their analogue in a 6d or 4d F-theory compactification on an elliptic fibration $ \hat X_3$ or $ \hat X_4$, respectively.
The cancellation of all gauge and mixed gauge-gravitational anomalies in both these classes of theories is captured by two relations, each valid in
$H^{4}(\hat X_3)$  or $H^{4}(\hat X_4)$, of the form
\bea \label{4d6d1}
 \sum_{{\bf R}, a}   \beta^a_\Gamma({\bf R}) \, \beta^a_\Lambda({\bf R}) \, \beta^a_\Sigma({\bf R})    S^a_{\bf R}  -  3 \, {\mathfrak F}_{(\Gamma} \cdot \pi^\ast \pi_\ast(\mathfrak{F}_\Lambda \cdot {\mathfrak F}_{\Sigma)})  &=& 0 \\
    \sum_{{\bf R}, a}   \beta^a_\Lambda({\bf R}) \,   S^a_{\bf R}   + 6   \, {\mathfrak F}_{\Lambda} \cdot c_1 &=& 0   \label{4d6d2} \,.
\eea
These two homological relations have been shown in \cite{Bies:2017abs} to be  equivalent to the intersection theoretic identities derived from the requirement of gauge and mixed gauge-gravitational anomaly cancellation in 6d \cite{Park:2011ji} and 4d \cite{Cvetic:2012xn} F-theory vacua. 
In addition the cancellation of purely gravitational anomalies in 6d F-theory vacua poses an extra constraint on the geometry of $\hat X_3$,
which has no direct counterpart in 4d.\footnote{\label{footnote6d}This relation is given, for example, as equation (3.8) in \cite{Park:2011ji}, and proven generally in \cite{Grassi:2000we}. } Interestingly enough, however, apart from this latter point anomaly cancellation in 6d and 4d F-theory vacua is based on the same type of homological relations.

While a general proof of these relations from first principles, and without relying on anomaly cancellation, is not yet available in the literature, these relations can be verified in explicit examples.\footnote{On the other hand, \cite{Corvilain:2017luj} proves anomaly cancellation in 4d F-theory vacua by comparison with the dual M-theory. Combined with the above statement this is a physics proof of  (\ref{4d6d1}) and (\ref{4d6d2}) on elliptic Calabi-Yau 4-folds.} The details of such a verification appear to be completely independent of the choice of base of the elliptic fibration, including its dimension \cite{Bies:2017abs}.
This raises the question if the same type of relations also holds on elliptically fibred Calabi-Yau 5-folds and if they play any role in anomaly cancellation in the associated 2d (0,2) theories.

The situation in compactifications to two dimensions looks rather more involved at first sight: 
As we have shown in section \ref{sec_AnomalyEqu5folds}, there are two types of independent anomaly relations, (\ref{gaugeanomalyFtheory}), associated with the cancellation of  the gauge anomaly, and another two, (\ref{gravitationalanomalies1}), for the pure gravitational anomaly.
We will now see that the flux dependent part of these anomaly relations, (\ref{gauge-flux1})   and (\ref{gravflux1}), is in fact closely related in form to (\ref{4d6d1}) and (\ref{4d6d2}).

Consider first relation (\ref{gauge-flux1}) for the cancellation of the flux dependent part of the 2d gauge anomalies,
\be
\label{GS1-v2}
\begin{split}
&  \sum_{{\bf R},a}   \beta^a_\Lambda({\bf R}) \beta^a_\Sigma({\bf R})   \pi_\ast (G_4 \cdot S^a_{\bf R})  \cdot_{C_{\bf R}} \pi_\ast (G_4 \cdot S^a_{\bf R})    \\
& = \pi_\ast (G_4 \cdot G_4) \cdot_{B_4}   \pi_\ast (\mathfrak{F}_\Lambda \cdot \mathfrak{F}_\Sigma) +     \pi_\ast (G_4 \cdot \mathfrak{F}_\Sigma) \cdot_{B_4}   \pi_\ast (G_4 \cdot \mathfrak{F}_\Lambda ) {+  \pi_\ast (G_4 \cdot \mathfrak{F}_\Lambda) \cdot_{B_4}   \pi_\ast (\mathfrak{F}_\Sigma \cdot G_4) \,.
}
\end{split}
\ee
A priori (\ref{GS1-v2}) holds for every transversal flux $G_4$, i.e. for every element $G_4 \in H^{2,2}(\hat X_5)$ satisfying (\ref{transverse1}), including potentially non gauge invariant fluxes. 
Our first observation is that this relation can be generalized to
\be
\label{GS1-v3}
\begin{split}
&  \sum_{{\bf R},a}   \beta^a_\Lambda({\bf R}) \beta^a_\Sigma({\bf R})   \pi_\ast (G^{(1)}_4 \cdot S^a_{\bf R})  \cdot_{C_{\bf R}} \pi_\ast (G^{(2)}_4 \cdot S^a_{\bf R})    \\
& = \pi_\ast (G^{(1)}_4 \cdot G^{(2)}_4) \cdot_{B_4}   \pi_\ast (\mathfrak{F}_\Lambda \cdot \mathfrak{F}_\Sigma) +     \pi_\ast (G^{(1)}_4 \cdot \mathfrak{F}_\Sigma) \cdot_{B_4}   \pi_\ast (G_4^{(2)} \cdot \mathfrak{F}_\Lambda ) {+  \pi_\ast (G_4^{(1)} \cdot \mathfrak{F}_\Lambda) \cdot_{B_4}   \pi_\ast (\mathfrak{F}_\Sigma \cdot G^{(2)}_4) }
\end{split}
\ee
valid for all transversal fluxes $G_4^{(1)}$ and $G_4^{(2)}$:
To see this, insert the ansatz  $G_4 = G_4^{(1)} + G_4^{(2)}$    into (\ref{GS1-v2}).  This gives three types of contributions, one depending quadratically on $G_4^{(1)}$ and on $G_4^{(2)}$, respectively, and a cross-term involving  $G_4^{(1)}$ and $G_4^{(2)}$. Since the quadratic terms vanish by themselves thanks to (\ref{GS1-v2}),  this is enough to establish the more general relation (\ref{GS1-v3}).

Let us now specialise one of the fluxes appearing in (\ref{GS1-v3}) to 
\bea \label{G41ansatz}
G_4^{(1)} = \pi^*D \cdot \mathfrak{F}_\Gamma \qquad {\rm with} \quad  D \in H^{1,1}(B_4)
\eea
and analyze the resulting identity further by repeatedly using the projection formulae
\bea
\pi_\ast (\pi^*A \cdot_{\hat X_5} B) &=& A \cdot_{B_4} \pi_\ast(B) \\
\pi_\ast(E) \cdot_{B_4} F &=& E \cdot_{\hat X_5} \pi^*(F) 
\eea
for suitable cohomology classes on $B_4$ and $\hat X_5$.
In the sequel, unless specified explicitly, the symbol $\cdot$ denotes the intersection product on $\hat X_5$. Then with (\ref{G41ansatz}) the first term on the RHS takes the form
\bea
\pi_\ast (G^{(1)}_4 \cdot G^{(2)}_4) \cdot_{B_4}   \pi_\ast ({\mathfrak F}_\Lambda \cdot \mathfrak{F}_\Sigma) &=& \left(D \cdot_{B_4} \pi_\ast({\mathfrak F}_\Gamma \cdot G_4^{(2)})\right) \cdot_{B_4} \pi_\ast(\mathfrak{F}_\Lambda \cdot \mathfrak{F}_\Sigma)  \\
&=& G_4^{(2)}  \cdot {\mathfrak F}_\Gamma \cdot \pi^\ast ( D \cdot_{B_4} \pi_\ast(\mathfrak{F}_\Lambda \cdot \mathfrak{F}_\Sigma) )  \\
&=&   \pi^\ast D \cdot G_4^{(2)}  \cdot {\mathfrak F}_\Gamma \cdot \pi^\ast \pi_\ast(\mathfrak{F}_\Lambda \cdot \mathfrak{F}_\Sigma)    \,.
\eea
Similar manipulations for the remaining two other terms on the RHS of (\ref{GS1-v2}) yield
\bea
{\rm RHS \,\,  of  \, \, (\ref{GS1-v2})} =  3 \,  \pi^*D \cdot G_4^{(2)} \cdot {\mathfrak F}_{(\Gamma} \cdot \pi^\ast \pi_\ast(\mathfrak{F}_\Lambda \cdot F_{\Sigma)}) \,.
\eea
As for the LHS, observe that 
\bea \label{G41SaR}
\pi_\ast(G_4^{(1)} \cdot S^a_{\bf R} ) =  \pi_\ast( \pi^\ast D \cdot {\mathfrak F}_\Gamma   \cdot S^a_{\bf R} ) = \beta^a_\Gamma({\bf R})  \left( D \cdot_{B_4} C_{\bf R} \right) \,.
\eea
Here we are using that in expressions of this form, the intersection of the divisor ${\mathfrak F}_\Gamma$ with the matter 3-cycle $S^a_{\bf R}$ in the fibre reproduces the charge $\beta^a_\Gamma$ of the associated state with respect to $U(1)_\Gamma$. As explained  around (\ref{c1LR}), the expression on the right of  (\ref{G41SaR}) is the first Chern class of the line bundle induced by the specific flux $G_4^{(1)}$ to which the matter states on $C_{\bf R}$ couple. For the special choice (\ref{G41ansatz}) this line bundle is the pullback of a line bundle from $B_4$.
With this understanding, the intersection product appearing on the LHS can be further simplified as
\bea
 \pi_\ast (G^{(1)}_4 \cdot S^a_{\bf R})  \cdot_{C_{\bf R}} \pi_\ast (G^{(2)}_4 \cdot S^a_{\bf R})  = \beta^a_\Gamma({\bf R})  \, \pi^*D \cdot G_4^{(2)} \cdot S^a_{\bf R} \,.
\eea
Altogether we have thus evaluated (\ref{GS1-v3}), for the special choice  (\ref{G41ansatz}), to
\bea
 \pi^\ast D  \cdot   G_4^{(2)} \cdot        \left( \sum_{{\bf R}, a}   \beta^a_\Gamma({\bf R}) \, \beta^a_\Lambda({\bf R}) \, \beta^a_\Sigma({\bf R})  \, S^a_{\bf R}  -  3 \,   {\mathfrak F}_{(\Gamma} \cdot \pi^\ast \pi_\ast(\mathfrak{F}_\Lambda \cdot {\mathfrak F}_{\Sigma)})    \right) = 0 \,.
 \eea
 Repeating the same steps for the flux dependent gravitational anomaly relation (\ref{gravflux1}) leads to 
 \bea
 \pi^\ast D  \cdot   G_4^{(2)} \cdot        \left(       \sum_{{\bf R}, a}   \beta^a_\Lambda({\bf R}) \,   S^a_{\bf R}   + 6   \, {\mathfrak F}_{\Lambda} \cdot c_1           \right) = 0 \,.
 \eea
 
The terms in brackets are identical in form with the linear combinations of 4-form classes which are guaranteed to vanish on an elliptically fibered Calabi-Yau 3-fold and 4-fold by anomaly cancellation according to (\ref{4d6d1}) and (\ref{4d6d2}). We conclude that \emph{if} the relations (\ref{4d6d1}) and (\ref{4d6d2}) hold also within $H^4(\hat X_5)$, as suggested by the results of \cite{Bies:2017abs}, this implies cancellation of the flux dependent part of the anomalies in 2d F-theory vacua for the special choice of flux (\ref{G41ansatz}). 
For more general fluxes, however, the constraints imposed on anomaly cancellation on a Calabi-Yau 5-fold seem to be stronger. In particular, 
a direct comparison with (\ref{4d6d1}) and (\ref{4d6d2}) is made difficult by the fact that 
(\ref{gauge-flux1})   and (\ref{gravflux1}) are quadratic in fluxes and a priori involve the intersection product on the matter loci $C_{\bf R}$, not on $B_4$. For general $G_4$ backgrounds, this makes a difference, as we have seen in section \ref{subsec_fluxdepan}. Furthermore, anomaly cancellation in 2d predicts the flux independent relations (\ref{gauge-geom1})   and (\ref{gravgeom1}). Condition (\ref{gravgeom1}) can be viewed as analogous, though very different in form, to the geometric condition on cancellation of the purely gravitational anomalies in 6d referred to in footnote \ref{footnote6d}.
It would be very interesting to investigate if a deconstruction of the topological invariants appearing in (\ref{gauge-geom1})   and (\ref{gravgeom1}), similar to the procedure applied for the Euler characteristic on Calabi-Yau 3-folds in \cite{Grassi:2000we,Grassi:2011hq}, can lead to a geometric proof of these identities.

\section{Conclusions and Outlook}

In this work we have provided closed expressions for the gravitational and gauge anomalies in 2d $N=(0,2)$ compactifications of F-theory on elliptically fibered Calabi-Yau 5-folds.
In particular, we have derived the Green-Schwarz counterterms for the cancellation of abelian gauge anomalies. The Green-Schwarz mechanism operates in a manner very similar to its 6d $N=(1,0)$ cousin: Dimensional reduction of the self-dual Type IIB 4-form results in real chiral scalar fields whose axionic shift symmetry is gauged and whose Chern-Simons type couplings hence become anomalous. 
We have uplifted our results for the gauging and the couplings to an expression valid in the most general context of F-theory on elliptically fibered Calabi-Yau 5-folds.
Anomaly cancellation in the 2d $(0,2)$ supergravity is then equivalent to (\ref{gaugeanomalyFtheory}) for the gauge and  (\ref{gravitationalanomalies1}) for the gravitational part. Each equation splits into a purely geometric and a flux dependent identity. These must hold separately on every elliptic Calabi-Yau 5-fold and for every consistent background of $G_4$ fluxes. We have verified this explicitly in a family of fibrations and for all vertical gauge fluxes thereon.

It is instructive to compare these 2d anomaly cancellation conditions to their analogue in 6d and 4d F-theory vacua in the form put forward in \cite{Park:2011ji} and \cite{Cvetic:2012xn}, respectively. 
The structure of anomalies as such becomes more and more constraining in higher-dimensional field theories.
At the same time the engineering of the quantum field theory in terms of the  internal geometry becomes more intricate as the dimension of the compactification space increases, and hence the number of large spacetime dimensions decreases. 
Correspondingly, the topological identities governing anomaly cancellation on elliptic 5-folds contain considerably more structure compared to their analogues in 4d and 6d F-theory compactifications. For once, the anomaly relations in 6d $N=(1,0)$ F-theory vacua are only sensitive to the topology of the elliptic fibration, while in 4d $N=1$ theories they are linearly dependent on a gauge flux. In 2d $N=(0,2)$ F-theories, both a purely topological and a flux dependent contribution arises. The latter is, in fact, quadratic in the gauge background. 

Despite differences in structure, the 6d and 4d gauge anomaly relations of \cite{Park:2011ji} and \cite{Cvetic:2012xn} can be reduced to one single identity \cite{Bies:2017abs}, valid in the cohomology ring $H^{2,2}(\hat X_{n})$ of an elliptically fibered Calabi-Yau $n$-fold, with $n=3$ and $4$, respectively. The same is true for their mixed gauge-gravitational counterparts. One motivation for the present work was to investigate these universal identities, (\ref{4d6d1}) and (\ref{4d6d2}), with respect to anomaly cancellation in 2d F-theories. The flux-dependent parts of (\ref{gaugeanomalyFtheory})  and  (\ref{gravitationalanomalies1})  exhibit striking similarities to (\ref{4d6d1}) and (\ref{4d6d2}). We have shown that if the 6d and 4d universal relations hold also in the cohomology ring of an elliptic 5-fold, as suggested by the examples studied in \cite{Bies:2017abs}, they imply the flux dependent anomaly relations at least for the subset of gauge backgrounds associated with massless $U(1)$ gauge groups. It would be very interesting to study further if also the converse is true, i.e. if the 2d relations allow us to establish a relation in the cohomlogy ring of elliptic 5-folds governing the 4d and 6d anomalies as well. 
The flux-independent anomaly relations, on the other hand, seem not to be related in a straightforward manner to the structure of anomalies in higher dimensions. 
In fact, already in 6d $N=(1,0)$ F-theory vacua, cancellation of the purely gravitational anomalies implies another topological identity with no counterpart in 4d. This relation has been proven quite generally in \cite{Grassi:2000we} using a deconstruction of the Euler characteristic of elliptic 3-folds. It would be worthwhile exploring if a similar proof is possible on Calabi-Yau 5-folds.

The structure of anomalies in 6d and 4d F-theory vacua is closely related to the Chern-Simons terms in the dual M-theory in five \cite{Intriligator:1997pq,Bonetti:2011mw,Bonetti:2013cza} and three dimensions \cite{Aharony:1997bx,Grimm:2011fx,Cvetic:2012xn}. In \cite{Corvilain:2017luj} this reasoning has lead to a proof of anomaly cancellation in 4d $N=1$ vacua obtained as F-theory on an elliptic Calabi-Yau 4-fold. It would be very interesting to extend such reasoning also to the 2d case. The Chern-Simons terms in the dual 1d $N=2$ Super-Quantum-Mechanics have been analyzed in \cite{Schafer-Nameki:2016cfr} and expressed geometrically in terms of data of the Calabi-Yau 5-fold. 
As expected, the similarities between the resulting identities such as (10.8) in \cite{Schafer-Nameki:2016cfr} and the 2d anomaly conditions are striking.

At a more technical level, the expressions for the anomalies presented in this work are valid under the assumption that the loci on the base hosting massless matter are smooth. 
Quite frequently, this assumption is violated, and an application of the usual index theorems requires a normalization of the singular loci \cite{Schafer-Nameki:2016cfr}. We leave it for future investigations to establish the anomaly relations in such more general situations. Likewise, in the presence of $\mathbb Q$-factorial terminal singularities in the fiber the precise counting of uncharged massless states in terms of topological invariants will change. In 6d F-theory vacua, this leads to a modification of the condition for cancellation of the gravitational anomaly \cite{Arras:2016evy,AGTW}, and similar effects are expected to play a role in 2d models.

Our focus in this work has been on the implications of anomaly cancellation rather than on the structure of the effective 2d $N = (0,2)$ supergravity per se. 
The axionic gaugings induced by the flux background, as derived in this context, give rise to a K\"ahler moduli dependent D-term, as noted already in \cite{Schafer-Nameki:2016cfr}.
What remains to be clarified is a careful definition of the chiral variables in the supergravity sector and a comparison of the Green-Schwarz action to the superspace formulation 
put forward in 2d (0,2) gauge theories in \cite{Adams:2006kb,Quigley:2011pv,Blaszczyk:2011ib}. This will also determine the correct normalization of the D-term.
At the level of the supersymmetry conditions induced by the flux, we have made, in passing, an interesting observation: Extrapolating from the situation on Calabi-Yau 4-folds we expect the existence of $G_4$ backgrounds which are not automatically of $(2,2)$ Hodge type and would hence break supersymmetry \cite{Haupt:2008nu}. More precisely, whenever $H^{2,2}(\hat X_5)$ contains $(2,2)$ forms which are not products of $(1,1)$ forms, it is expected that the Hodge type of a 4-form varies over the complex structure moduli space. This would constrain some of the complex structure of the 5-fold \cite{Haupt:2008nu}. 
This makes it tempting to speculate that the contribution of the supergravity sector to the purely gravitational anomaly should change compared to a background without flux.
At the same time, the flux dependent contribution to the D3-brane tadpole modifies the class of the D3-branes in the background and therefore also the anomaly contribution from the sector of 3-7 string modes. For consistency, both effects have to cancel each other, which is in principle possible due to the opposite chirality of the fields involved.
In this sense the net effect of complex moduli stabilization would be topological, in stark contrast to the situation in 4d $N=1$ compactifications. More work on elliptically fibered 5-folds is needed to flesh out the details behind this phenomenon.

\noindent {\bf Acknowledgements} 

We thank Seung-Joo Lee and Diego Regalado for important discussions, and Martin Bies, Antonella Grassi, Craig Lawrie, Christoph Mayrhofer and Sakura Sch{\"a}fer-Nameki in addition for collaboration on related topics. The work of T.W. and F.X. is partially supported by DFG Transregio TR33 'The Dark Universe' and  by DFG under GK 'Particle Physics Beyond the Standard Model'.



\appendix
\section{Conventions}\label{convention}

In this appendix we collect our conventions for the technical computations in this paper.

\subsection{Local Anomaly}

Our conventions for the anomaly polynomial mostly  follow 
\cite{Bilal:2008qx}. 
Consider a quantum field theory in $D=2r$-dimensional Minkowski space $M_{2r}$ with quantum effective action $S[A]$, where $A_\alpha$ is the connection associated with a local symmetry of $S$ with gauge parameter $\epsilon^\alpha$.
The anomaly $\mathfrak{A}_\alpha$  is defined as the gauge variation 
\be
\delta_\epsilon S[A]= \int_{M_{2r}} \epsilon^\alpha \mathfrak{A}_\alpha  \,.  
\ee
 It is expressible as
\be
\int_{M_{2r}} \epsilon^\alpha \mathfrak{A}_\alpha=2\pi \int_{M_{2r}} I_{2r}^{(1)}(\epsilon) \,,
\ee
where the $2r$-form $I_{2r}^{(1)}(\epsilon)$ is related to  $(2r+2)$-form $I_{2r+2}$ via  the Stora-Zumino descent relations 
\be
I_{2r+2}=dI_{2r+1}, \qquad \delta_\epsilon I_{2r+1}=d I_{2r}^{(1)}(\epsilon) \,.
\ee
In our sign conventions, the anomaly polynomial $I_{2r+2}$ of a complex chiral Weyl fermion in representation ${\bf R}$ takes the form
\be
\label{anomalypoly}
I_{s=1/2}({\bf R})|_{2r +2} = - {\rm tr}_{\bf R} e^{-F}\hat A({\rm T}) |_{2r +2} \,.
\ee
Here $F$ is the hermitian field strength associated with the gauge potential $A$ and ${\rm T}$ denotes the tangent bundle to spacetime. Its curvature 2-form $R$ is the curvature associated with the spin connection.
Furthermore, in $2r = 4k +2$ dimensions, a self-dual $r$-tensor contributes to the gravitational anomalies with
\bea \label{anomalypolyb}
I_{\rm s.d.}|_{2r + 2}= - \frac18L({\rm T})|_{2r+2} \,.
\eea
The A-roof genus and the Hirzebruch L-genus above can be expressed as 
\be
\ba \label{AroofL}
\hat A({\rm T}) &= 1 - \frac{1}{24}\,  p_1({\rm T}) + \ldots = 1 - \frac{1}{24} (c_1^2({\rm T}) - 2 c_2({\rm T})) + \ldots \cr
L({\rm T}) &= 1 + \frac{1}{3} \, p_1({\rm T}) + \ldots = 1 + \frac{1}{3}(c_1^2({\rm T}) - 2 c_2({\rm T})) + \ldots \,.
\ea
\ee
We will oftentimes write the first Pontrjagin class of the tangent bundle as
\bea
p_1({\rm T}) = - \frac{1}{2} {\rm tr} R \wedge R \,.
\eea
Note that we have included an overall minus sign in (\ref{anomalypoly}) and (\ref{anomalypolyb}) compared to the conventions used in \cite{Bilal:2008qx}.
The reason is that in the quantum field theory we are analyzing the chiral fermion fields arise as the zero-modes of strings on the worldvolume of 7- and 3-branes.
The  anomalies induced by these modes on the brane worldvolume must be cancelled via an anomaly inflow mechanism by the anomalous Chern-Simons action of the branes. This relates the sign of the 1-loop anomalies to the sign conventions used for the Chern-Simons brane actions.
 As we will discuss below, the sign of the 7-brane Chern-Simons action is fixed as in (\ref{CScouplingsFtheory}) by the convention that the 7-brane couples magnetically to the axio-dilaton, which is usually defined in F-theory as $\tau = C_0 + i e^{-\phi}$ (rather than $-C_0 + i e^{\phi}$).
The sign chosen in (\ref{CScouplingsFtheory}) conforms with this convention. In order for the anomalies of chiral fermions in the worldvolume of a D7-brane to be cancelled by anomaly inflow, we must then adopt the convention (\ref{anomalypoly}).

\subsection{Type IIB 10D supergravity and brane Chern-Simons actions} \label{app_conventions}

The bosonic part of the 10d Type IIB supergravity pseudo-action in its democratic form is given by 
\be
\label{10d}
\ba
S_{\rm IIB} &=&  2 \pi \left(\int d^{10}x\ e^{-2 \phi} ( \sqrt{-g} R + 4 \partial_M \phi \, \partial^M \phi) -\frac{1}{2} \int e^{-2 \phi}  H_3 \wedge \ast H_3  \right. \cr
&&\left. - \frac{1}{4}   \sum_{p=0}^4  \int   F_{2p+1}  \wedge \ast F_{2p+1} - \frac{1}{2} \int C_4 \wedge H_3 \wedge F_3   \right).
\ea
\ee
Here we are working in conventions where the string length $\ell_s = 2 \pi \sqrt{\alpha'} \equiv 1$ and the field strengths are defined as
\be
\ba
& H_3 = dB_2, \quad F_1 = dC_0, \quad F_3 = dC_2 - C_0 \, dB_2, \cr 
& F_5 = dC_4 - \frac12 C_2 \wedge dB_2 + \frac12 B_2 \wedge dC_2,
\ea
\ee
together with the duality relations $F_9 = \ast F_1$, $F_7 = - \ast F_3$, $F_5 = \ast F_5$, which hold at the level of equations of motion.

The Chern-Simons action for the D7-branes and the O7-plane takes the form
\be \label{CScouplingsFtheory}
\ba
S^{D7} &= -\frac{2 \pi}{2} \int_{D_7} {\rm Tr} \, e^{i\cal F}  \, \sum_{2p} C_{2p} \, \sqrt{  \frac{ \hat A({\rm T}D_7)}{\hat A({\rm N}D_7)}} \cr
S^{O7} &=  \frac{16 \pi}{2} \int_{O_7} \sum_{2p} C_{2p} \, \sqrt{ \frac{L(\frac{1}{4} {\rm T}O_7)}{L(\frac{1}{4} {\rm N}O_7)}} \,.
\ea
\ee
Since we are working in the democratic formulation, where each RR gauge potential is accompanied by its magnetic dual, the Chern-Simons action has to include a factor of $\frac{1}{2}$ \cite{Cheung:1997az}, which we are making manifest in (\ref{CScouplingsFtheory}). This factor is crucial in order to obtain the correctly normalized anomaly inflow terms, and, as we find in the main text, also to reproduce the correctly normalised Green-Schwarz counterterms.
As stressed already, the minus sign in front of the Chern-Simons action of the D7-branes ensures that in the above conventions for the supergravity fields the D7-brane couples magnetically to the  
axio-dilaton $\tau=C_0+i e^{-\phi}$.
Note furthermore that we are writing the brane action in terms of ${\rm Tr} = \frac{1}{\lambda}{\rm tr}_{\bf fund}$, where the Dynkin index $\lambda$ is given in \autoref{Tab_Lambda}. 
Finally, ${\rm T}D_7$ and ${\rm N}D_7$ denote the tangent and normal space to the 7-brane along $D_7$, and similarly for the O7-plane.
The Chern-Simons action for a D3-brane carries a relative sign compared to the 7-brane action, 
\be
S^{D3} = \frac{2 \pi}{2} \int_{D3} {\rm Tr} \, e^{i\cal F}  \, \sum_{2p} C_{2p} \, \sqrt{  \frac{ \hat A({\rm T}D3)}{\hat A({\rm N}D3)}} \,.
\ee

The gauge invariant field strength $\mathcal{F}$ above is defined as
\bqa
\mathcal{F}=i(\mathbf{ F}+2\pi\phi^\ast B_2\mathbb{I}) \,.
\eqa
Compared to expressions oftentimes used in the literature we have absorbed a factor of $\frac1{2\pi}$ in the definition of $\mathcal{F}$. The NS-NS two-form field  $B_2$ is pulled back to the brane via  $\phi^\ast$. We will always set $B_2=0$ in this article, but one should bear in mind that it appears in various consistency conditions as  detailed e.g. in \cite{Blumenhagen:2008zz}. We will sometimes decompose
\be
\label{gaugefield}
\mathbf{ F}=F+\bar{F}
\ee
so that  $F$ denotes the gauge invariant field strength of the gauge field in non-compact flat space while $\bar{F}$ stands for the internal flux background. Note that it is the hermitian field strength $F$ which appears in the anomaly polynomial \eqref{anomalypoly}.
Finally, the curvature terms in the above Chern-Simons actions enjoy the expansion 
\be
 \sqrt{  \frac{ \hat A({\rm T}D_7)}{\hat A({\rm N}D_7)}} = 1 + \frac{1}{24} c_2(D_7) + \ldots,  \qquad \sqrt{ \frac{L(\frac{1}{4} {\rm T}O_7)}{L(\frac{1}{4} {\rm N}O_7)}} = 1 - \frac{1}{48}  c_2(O_7) + \ldots \,.
 \ee
Here we have used the definitions (\ref{AroofL}) together with the fact that $c_1({\rm T}D) = - c_1({\rm N}D)$ by adjunction on the Calabi-Yau space on which we compactify the Type IIB theory

 \subsection{Type IIB orientifold compactification with 7-branes}
 
In a Type IIB orientifold compactification on a Calabi-Yau 4-fold $X_4$, the orientifold projection $\Omega (-1)^{F_L} \sigma$ acts as in the more familiar case of compactification on a 3-fold, as summarized e.g. in \cite{Grimm:2004uq}. 
In particular, the  $p$-form fields transform under the combined action of worldsheet parity $\Omega$ and left-moving femrion number $(-1)^{F_L}$ as
 \be \label{parity}
\Omega (-1)^{F_L} : \qquad (C_0, B_2, C_2, C_4,C_6) \ \rightarrow \ (C_0, -B_2, -C_2, C_4,-C_6)\,.
\ee
The holomorphic involution $\sigma$ acts only on the internal space $X_4$ such that the K\"ahler form $J$ and the holomorphic top-form $\Omega_{4,0}$ transform as
\be
\sigma: \qquad J \rightarrow J\,, \qquad \Omega_{4,0} \rightarrow - \Omega_{4,0} \,.
\ee
The cohomology groups $H^{(p,q)}(X_4)$ split into two eigenspaces $H^{(p,q)}(X_4)=H^{(p,q)}_+(X_4)\bigoplus H^{(p,q)}_-(X_4)$ under the action of $\sigma$.
In performing the dimensional reduction, the orientifold even and odd form fields are expanded along a basis of the invariant and anti-invariant cohomology groups.

Under the orientifold action the field strength on each brane is mapped to its cousin on the orientifold image brane
 \bea
{\cal F}_i \rightarrow    {\cal F}'_i =  - \sigma^* {\cal F}_i \,,
\eea
where the minus sign is due to the worldsheet parity action.

\section{Anomalies and Green-Schwarz term in Type IIB orientifolds} \label{app_AnomaliesIIB}

In this appendix  we verify our intermediate results \eqref{GSterm} for the Green-Schwarz terms in Type IIB orientifolds. Together with our confirmation of the final F-theoretic expressions in the explicit example of section \ref{sec_ExampleSU5U1}, this also supports our rules explained in (\ref{sec_Ftheorylift}) for the correct uplift to F-theory.

The setup we analyze is identical to the one in appendix C.2 of \cite{Schafer-Nameki:2016cfr}, which we now briefly summarize. 
Consider a Type IIB orientifold on a general Calabi-Yau 4-fold $X_4$ with gauge group $(SU(n)\times U(1)_a) \times U(1)_b$. The brane configuration consists of $n$ 7-branes wrapping a divisor $W$ and one extra D7-brane along the divisor $V$, each accompanied by their orientifold images wrapped along $W'$ and $V'$, respectively. We assume that all brane divisors are smooth. 
In order to cancel the D7-tadpole, it is required that 
\be
n([W]+[W'])+([V]+[V'])=8[O7] \,.
\ee
The D3-tadpole cancellation condition fully determines the spacetime-filling D3-brane system wrapped along a total curve class $[C]$ plus orientifold image brane $[C']$ as 
\bea \label{D3tadpoleIIB-app}
[C] &=& \frac{n}{24}[W]\cdot c_2(W)+\frac1{12} [O7]\cdot c_2(O7) + n \, {\rm ch}_2(L_W)\cdot [W]   + {\rm ch}_2(L_V)\cdot [V] \\
{} [C'] &=& \frac{n}{24} [W']\cdot c_2(W')+\frac1{12}  [O7]\cdot c_2(O7) +n \, {\rm ch}_2(L_W')\cdot [W'] + {\rm ch}_2(L_V')\cdot [V'] \,.
\eea
Here $L_W$ and $L_V$ denote  line bundles on $W$ and $V$ whose structure groups are identified with $U(1)_a$ and $U(1)_b$, respectively.

For simplicity, we require 
\be
[V]=[V'], \qquad [W]=[W']
\ee
to prevent the gauge potentials associated with $U(1)_a$ and $U(1)_b$ from acquiring a mass, in absence of flux, via the geometric St\"uckelberg mechanism.\footnote{Otherwise, a D5-bane tadpole cancellation must be imposed on the gauge background.} 
We simplify the calculation of the $U(1)_a$ anomaly contribution further by assuming  
\be \label{WW'constraint}
[W]\cdot[W']=[W]\cdot[O7]   \,.
\ee
This implies that there exists no intersection locus of $W$ and $W'$ away from the $O7$-plane, which would carry matter in the symmetric representation of $SU(n)$.
This would lead to extra complications in the computation of the chiral spectrum, which we avoid by requiring (\ref{WW'constraint}).
 For the same reason we make the simplifying assumption that  
\be \label{VV'constraint}
 [V] \cdot [V']=[V] \cdot [O7] \,.
 \ee
None of these assumptions is essential, but dropping them would require some modifications of the anomaly computation.

We are now in a position to determine the contribution to the  $U(1)_a - U(1)_a$ and the  $U(1)_b - U(1)_b$  anomaly due to the chiral matter states.
Since our primary interest here is to check the Green-Schwarz counterterm \eqref{GSterm} and its normalization relative to the 1-loop anomalies, it suffices to
 focus on the flux-dependent contribution of these states. The chiral spectrum from the D7-D7 brane sector and the flux dependent part of its contribution to the anomalies
 are listed in table \ref{7-7}, and similarly for the 3-7 sector in table \ref{3-7}. Note that we have omitted matter in the adjoint representation, which is not charged under $U(1)_a$ and $U(1)_b$.  We adapt the convention  \eqref{anomalypoly} for the anomaly polynomial so that there is overall factor of $-1$ in front of every term in \autoref{7-7}, while in \autoref{3-7} we have taken into account the anti-chiral nature of the 3-7 matter, which hence contributes with a $+1$.
 Merely to save some writing, we have assumed, in the column containing the $U(1)_a^2$ anomalies, that $L_V = 0$, and similarly in the column containing the $U(1)^2_b$ anomalies that $L_W = 0$. Furthermore, with our assumption (\ref{WW'constraint}) all matter on $W \cap W'$ transforms in the anti-symmetric representation of $U(n)$, while due to (\ref{VV'constraint}) the states on $V \cap V'$ are all projected out (as there exists no anti-symmetric representation of $U(1)_b$). 
 The total anomaly from the 7-7 sector is then obtained by summing over all states in table \autoref{7-7} and dividing the final result by two.
The division by two is due to the orientifold quotient.
 \autoref{7-7} contains sectors in this upstairs picture which are pairwise identified under the involution. To offset for this overall factor of  $\frac{1}{2}$ in the invariant sector $W \cap W'$ we are including a factor of $2$ for these states   in \autoref{7-7}.

\begin{table}[t]
\begin{center}
\begin{tabular}{c|c|c|c|}
 Locus & Representation & $U(1)^2_a$ anomaly  ($c_1(L_V) = 0$) &$U(1)^2_b$ anomaly  ($c_1(L_W) = 0$) \\
  & of $SU(n)_{q_a,q_b}$    & $=-\frac12 \int {\rm ch}_2(L)q_a^2{\rm dim}(R)$ &$=-\frac12 \int {\rm ch}_2(L)q_b^2{\rm dim}(R)$\\
\hline
$W\bigcap V$ & $\bar n_{(-1,1)}$ & $-\frac12  [W] \cdot [V]  \cdot   \frac12c_1^2(L_W) \, n   
  $&$-\frac12 [W] \cdot [V]  \cdot     \frac12c_1^2(L_V) \, n $  \\ 
$W\bigcap V'$ & $\bar n_{(-1,-1)}$  & $-\frac12    [W] \cdot [V'] \cdot  \frac12 c_1^2(L_W) \, n$ &$-\frac12  [W] \cdot [V']  \cdot  \frac12c_1^2(L_V) \, n$    \\ 
$W'\bigcap V'$ & $ n_{(1,-1)}$& $-\frac12    [W'] \cdot [V']  \cdot  \frac12c_1^2(L_W) \, n$&  $-\frac12   [W'] \cdot [V'] \cdot  \frac12c_1^2(L_V) \, n$\\
$W'\bigcap V$ & $ n_{(-1,-1)}$& $-\frac12   [W'] \cdot [V] \cdot    \frac12 c_1^2(L_W) \, n$ & $-\frac12   [W'] \cdot [V]  \cdot \frac12c_1^2(L_W) \, n$\\
$W'\bigcap W$ & $\frac12 n(n-1)_{(2,0)}$&  $ - 2\times \frac12  [W] \cdot [W'] \cdot \frac12c_1^2(L^2_w) \times 2^2\times \frac12 n(n-1)$ &0\\
\end{tabular}
\caption{Charged chiral matter from the 7-7 string sector and its anomaly contributions.}
\label{7-7}
\end{center}
\end{table}

From (\ref{D3tadpoleIIB-app}) we read off the flux-dependent term part of the 3-brane  class $[C]$, 
\be
[C]|_{\rm flux}=\frac12 n \, c_1^2(L_W)\cdot [W]+\frac12 c_1^2(L_V)\cdot [V] \,.
\ee

The 7-7 and 3-7 sector contribution to the $U(1)_a - U(1)_a$ anomaly is hence, for $c_1(L_V) = 0$ for simplicity,
\be
\left. I_4^{\rm 1-loop} \right |_{U(1)^2_a} = F_a^{\rm 2d} \wedge F_a^{\rm 2d}  \,   {\cal A}_a
\ee
with
\be
\ba
\mathcal{A}_a&=-\frac12\left( \big( 4\times \frac14 \big) c_1^2(L_W) \, n \, [W]\cdot [V]+ 4\times \frac14 \frac22 n(n-1) \, c_1^2(L^2_W) \, [W]\cdot 4[O7] \right.  \cr
 & \phantom{=}-  \left.4\times \frac12 n [W]\cdot \frac12 nc_1^2(L_W)\cdot [W] \right)\cr
&=-\frac12 [W] \cdot c_1^2(L_W) \cdot \left( n[V]+ 4 n^2 \,  [O7] -  4 n \, [O7] -n^2 \, [W] \right) \cr
&=- n^2  \,  [W]^2 \cdot c_1^2(L_W)  \,.
\ea
\ee
In the last line we have used that
\be
n[W]+[V]=4[O7], \qquad [W]\cdot [W]=[W]\cdot [O7] \,.
\ee
This 1-loop anomaly is precisely cancelled by the Green-Schwarz term contribution \eqref{GSterm} because the trace over the diagonal $U(1)_a \subset U(n)$ evaluates to
${\rm Tr}   \bar F_a={\rm tr} \mathbb{I}_n \, \bar F_a$ and hence
\be
\left. I_4^{\rm GS}\right |_{U(1)_a^2} = \frac{1}{4}  {\rm Tr}_a {\rm Tr}_a F_a^{\rm 2d} \wedge F_a^{\rm 2d}    \left(  4 \,  \bar F_a \cdot [W] \right) =  F_a^{\rm 2d} \wedge F_a^{\rm 2d}    \left( n^2 c_1(L_W) \cdot [W] \right) \,.
\ee

\begin{table}[t]
\begin{center}
\begin{tabular}{c|c|c|c|c|} Locus & Representation & $U(1)^2_a$  ($c_1(L_V) = 0$) & $U(1)_b$  ($c_1(L_W) =0$) \\
  & of $SU(n)_{q_a,q_b}$    & $=+\frac12 W\cdot C q^2{\rm dim}(R) $ &$=+\frac12 V\cdot C q^2{\rm dim}(R) $\\
\hline
$W\bigcap C$ & $\bar n_{(-1,1)}$ & $+\frac12 [W] \cdot [C]\times 1^2\times n $&0\\
$W\bigcap C'$ & $\bar n_{(-1,-1)}$  & $+\frac12 [W]\cdot [C]\times 1^2\times n $&0\\
$W'\bigcap C'$ & $ n_{(1,0)}$& $+\frac12 [W]\cdot [C]\times 1^2\times n $ &0\\
$W'\bigcap C$ & $ n_{(1,0)}$& $+\frac12 [W]\cdot [C]\times 1^2\times n $ &0\\
$V\bigcap C$ & $ 1_{(0,-1)}$& 0&$+\frac12 [V]\cdot [C]\times 1^2\times 1 $ \\
$V\bigcap C'$ & $ 1_{(0,-1)}$& 0&$+\frac12 [V]\cdot [C']\times 1^2\times 1 $ \\
$V'\bigcap C$ & $ 1_{(0,1)}$& 0&$+\frac12 [V']\cdot [C]\times 1^2\times 1 $ \\
$V'\bigcap C'$ & $ 1_{(0,1)}$& 0&$+\frac12 [V']\cdot [C']\times 1^2\times 1 $ \\
\end{tabular}
\caption{Charged chiral matter from the 3-7 string sector and its anomaly contributions.}    
\label{3-7}
\end{center}
\end{table}

Similarly, the 1-loop $U(1)^2_b$ anomaly induced by the chiral matter, for $c_1(L_W) = 0$, 
\be
\ba
\mathcal{A}_b&=-\frac12  \left( (4\times \frac14) \, c_1^2(L_V) \, n \, [W]\cdot [V]  -  4\times \frac12 [V] \cdot \frac12 \, c_1^2(L_V)\cdot [V] \right) \cr
&=-  [V]^2 \cdot c_1^2(L_V)
\ea
\ee
is correctly cancelled by  the GS term contribution \eqref{GSterm}.


\section{Chirality computation for matter surface flux} \label{app_chirality}

In this appendix we compute flux dependent part of the chiral index (\ref{chi5-2lambda}) induced for states in representation ${\bf 5}_{-2}$ by the gauge background $G_4^\lambda$ in the $ SU(5) \times U(1)_A$ model of section \ref{subsec_fluxdepan}.
The matter surface $C_{{\bf 5}_{-2}} \subset W \subset B_4$ is cut out by the locus $P \cap W$ on $B_4$ with 
\bea
P:=\{a_{1} a_{4,3} - a_{2,1} a_{3,2} =0\}\,.
\eea
The classes in which the Tate polynomials $a_{i,j}$ take their value are listed in (\ref{classtatecoefficient}).
As discussed in section \ref{subsec_fluxdepan}, 
our task amounts  to computing
\bea \label{intc125-2}
\frac{1}{2 }\int_{C_{{\bf 5}_{-2}}}   c_1^2(L_{{\bf 5}_{-2}}) =   \frac{\lambda^2}{50}    \int_{C_{{\bf 5}_{-2}}}  (   - 2 [Y_2] +  3 [Y_1] )^2 \,,
\eea
where $[Y_1]$ and $[Y_2]$ denote the classes of eponymous curves on the surface $C_{{\bf 5}_{-2}} \subset W \subset B_4$. 
These curves cannot be expressed as the complete intersection of the surface $C_{{\bf 5}_{-2}}$  with a divisor from $B_4$, but are defined by the complete intersection of 
7-brane divisor $W$ with two divisors on $B_4$. Concretely, from  (\ref{defYi}) we read off
\bea
Y_i =   A_i \cap B_i
\eea
with
\bea \label{AiBidef}
[A_1] = [A_2] =  [a_1]|_W,  \qquad B_1 = [a_{2,1}]|_W,  \qquad [B_2]=[a_{3,2}]|_W. 
\eea
We hence need to evaluate the intersections $\int_{C_{{\bf 5}_{-2}}}   [Y_i] \cdot [Y_j]$ for $i=1,2$.
The self-intersections of $Y_i$ on $C_{{\bf 5}_{-2}}$ are computed via
\bea \label{intyiyia}
\int_{C_{{\bf 5}_{-2}}}  [Y_i] \cdot [Y_i] = \int_{Y_i} [Y_i] =    \int_{Y_i} c_1(N_{Y_i \subset C_{{\bf 5}_{-2}}}) \,,
\eea
where the first Chern class of the normal bundle $N_{Y_i \subset C_{{\bf 5}_{-2}}}$ is computed via the normal bundle short exact sequence
\bea
0 \rightarrow N_{Y_i \subset C_{{\bf 5}_{-2}}}  \rightarrow N_{Y_i \subset W} \rightarrow N_{C_{{\bf 5}_{-2}}  \subset W} \rightarrow 0 \,.
\eea
The normal bundles are given as
\bea
N_{Y_i \subset W} &=& {\cal O}(A_i) \oplus {\cal O}(B_i)  \\
N_{C_{{\bf 5}_{-2}}  \subset W} &=& {\cal O}(P) \,,
\eea
where ${\cal O}(A_i)$   defines a line bundle of first Chern class $[A_i] |_ {C_{{\bf 5}_{-2}}}$ on ${C_{{\bf 5}_{-2}}}$    and  ${\cal O}(P)$ is a line bundle on $W$ of first Chern class $[P]|_W$.
This gives
\bea
c(N_{Y_i \subset C_{{\bf 5}_{-2}}}  ) &=& \left. \frac{c(N_{Y_i \subset W})}{c(N_{C_{{\bf 5}_{-2}}  \subset W})} \right|_{C_{{\bf 5}_{-2}}}= \left. \frac{1 + c_1 (N_{Y_i \subset W})}{ 1 + c_1(N_{C_{{\bf 5}_{-2}}  \subset W})} \right|_{C_{{\bf 5}_{-2}}}\\
&=& \left.(1 + [A_i] + [B_i] ) ( 1 - [P] + [P]^2 + \ldots) \right|_{C_{{\bf 5}_{-2}}}\,.
\eea
Collecting the terms of first order yields
\bea
c_1(N_{Y_i \subset C_{{\bf 5}_{-2}}}  )  = \left. (- [P]  + [A_i] + [B_i] ) \right|_{C_{{\bf 5}_{-2}}} \,.
\eea
The integral (\ref{intyiyia}) can now be expressed as an integral directly on $W$,
\bea
\int_{ Y_i}  c_1(N_{Y_i \subset C_{{\bf 5}_{-2}}}  )  = ([A_i] \cdot_{W} [B_i]) \cdot_{W} ( - [P] |_{W} + [A_i] + [B_i]) \,.
\eea 
Since all involved classes are defined on or can be extended to $B_4$, this evaluates to 
\bea
\int_{ Y_1}  [Y_1] =2 c_1 \cdot W \cdot (2 c_1-W) \cdot ( W- c_1),\qquad \int_{ Y_2}  [Y_2] =c_1 \cdot W \cdot (3 c_1-2 W) \cdot (W-c_1) \,,
\eea
in terms of the intersection product on $B_4$, where we are using (\ref{AiBidef})  and (\ref{classtatecoefficient}).

The remaining task is to compute the cross-term $\int_{C_{{\bf 5}_{-2}}}   [Y_1] \cdot [Y_2]$. 
We note that even though the curves $Y_i$ cannot individually be written as the complete intersection of a divisor with the divisor $P$ defining $C_{{\bf 5}_{-2}}$, the combination $Y_1 + Y_2$ is of this simpler form:
Indeed on $W$ we have that
\bea
Y_1 + Y_2 =  C_{{\bf 5}_{-2}} \cap C_{{\bf 10}_{1}} \,.
\eea
Since $C_{{\bf 10}_{1}}  = \{a_1 = 0 \}$ we can then write on $C_{{\bf 5}_{-2}}$ for $Y_1 + Y_2$
\bea
Y_1 + Y_2   = \{  a_1|_{C_{{\bf 5}_{-2}}} = 0 \}  \subset C_{{\bf 5}_{-2}} \,.
\eea
In particular, with $ [a_1] = c_1$,
\bea
\int_{C_{{\bf 5}_{-2}}} ([Y_1] + [Y_2])^2 = \int_{C_{{\bf 5}_{-2}}}  [a_1] \cdot [a_1] =  c_1^2 \cdot W \cdot (5c_1-3W) \,,
\eea
where the last intersection is taken on $B_4$. 
The idea is then to express the cross-term as 
\bea
\int_{C_{{\bf 5}_{-2}}} [Y_1] \cdot [Y_2] = \int_{C_{{\bf 5}_{-2}}} \frac{1}{2} \left(    ([Y_1]+ [Y_2])^2 - [Y_1]^2 - [Y_2]^2 \right) = c_1 \cdot  W \cdot  (6 c_1^2-7 c_1 W+2 W^2) \,.
\eea
Plugging everything into (\ref{intc125-2}) leads to the final result (\ref{chi5-2lambda}).



\newpage
\bibliography{papers}
\bibliographystyle{custom1}

\end{document}